\documentclass[12pt]{article}
\usepackage{graphicx}
\usepackage{pgfplots}
\usepackage{booktabs}
\usepackage{amsmath}
\usepackage{algorithm}
\usepackage{algorithmic}
\usepackage{amssymb}
\usepackage{amsfonts}
\usepackage{latexsym}
\usepackage{color}
\usepackage{subfigure}
\usepackage{latexsym,amsmath,amsfonts}
\usepackage{graphicx,epstopdf}
\usepackage{subfigure}
\usepackage{authblk}
\usepackage{caption}

\newcommand{\be}{\begin{equation}}
\newcommand{\en}{\end{equation}}
\newcommand{\bea}{\begin{eqnarray}}
\newcommand{\ena}{\end{eqnarray}}
\newcommand{\beano}{\begin{eqnarray*}}
\newcommand{\enano}{\end{eqnarray*}}
\newcommand{\bee}{\begin{enumerate}}
\newcommand{\ene}{\end{enumerate}}
\newcommand{\R}{{\cal R}}

\newcommand{\Hil}{{\cal H}}

\newcommand{\Pc}{{\cal P}}
\newcommand{\Sc}{{\cal S}}

\newcommand{\1}{1 \!\! 1}
\newtheorem{thm}{Theorem}

\newtheorem{prop}[thm]{Proposition}
\newtheorem{defn}[thm]{Definition}

\newtheorem{remark}{Remark}

\begin{document}

\title{$(H,\rho)$-induced dynamics \\and the quantum game of life}
\author{F. Bagarello$^{1,3}$, R. Di Salvo$^{2}$, F. Gargano$^1$, F. Oliveri$^{2}$\\
\small{
$^1$DEIM -  Universit\`a di Palermo,
Viale delle Scienze, I--90128  Palermo, Italy,}\\
\small{$^2$MIFT -- Universit\`a di Messina,
Viale F. Stagno d'Alcontres 31, I--98166 Messina, Italy}\\
\small{ $^3$I.N.F.N -  Sezione di Torino.}\\
\small{\emph{Email addresses:}\\
fabio.bagarello@unipa.it,
rosa.disalvo@unime.it,\\
francesco.gargano@unipa.it,
francesco.oliveri@unime.it}
}

\date{}
\maketitle
\begin{abstract}
We propose an extended version of quantum dynamics for a certain system $\Sc$, whose evolution is ruled by a Hamiltonian $H$, its initial conditions, and a suitable set $\rho$ of  {\em rules}, acting repeatedly on $\Sc$. The resulting dynamics is not necessarily periodic or quasi-periodic, as one could imagine for conservative systems with a finite number of degrees of freedom. In fact, it may have quite different behaviors depending on the explicit forms of $H$, $\rho$ as well as on the initial conditions. After a general discussion on this $(H,\rho)$-{\em induced dynamics}, we apply our general ideas to extend the classical game of life, and we analyze several aspects of this extension.
\end{abstract}


\section{Introduction}\label{sec:intro}

The  Game of Life (hereafter, \emph{GoL}) can be thought of as a sort of a dynamical system $\Sc$ in which
we are interested to the changes of the local densities of a given population $\Pc$ living in a lattice $\R$. In
the generic cell $C_j$ of $\R$, the density of the population changes according to what happens in the other cells
surrounding $C_j$ itself (typically, the eight surrounding cells characterizing the so-called Moore neighborhood);  in particular,
this change is driven by the sum of the densities of the populations in these other cells.
In other words, the  \emph{GoL} is a two-dimensional cellular automaton
in which each cell  at any time assumes only two possible values: 0 if the cell is in a dead state, 1 if the cell is alive.  At each  \emph{generation}, a given cell undergoes a transition according to
specific rules based on its own state and on the states of the surrounding cells.
More formally, we can write the \emph{GoL} as the  cellular automaton
\[
A_{GoL}=\{Z_{GoL}^2,\mathcal{N},\{0,1\},f\},
\]
where $Z_{GoL}$  is the set of all integers
such that $Z_{GoL}^2$ represents the two-dimensional array of the cellular space, $\mathcal{N}$ is the Moore neighborhood index, $\{0,1\}$ is the set of the possible states of a cell, and $f:\{0,1\}^{|\mathcal{N}|+1}\rightarrow \{0,1\}$ is the transition function defined as
 \begin{eqnarray}
 f(1,\{\alpha\})&=&1 \textrm{ if } \left(\left (\sum_{\alpha\in \mathcal{N}
    } \alpha=2\right) \vee  \left(\sum_{\alpha \in \mathcal{N}
      } \alpha=3\right)\right),\label{r1}\\
  f(1,\{\alpha\})&=&0 \textrm{ if } \left(\left (\sum_{\alpha \in \mathcal{N}
   } \alpha<2\right) \vee  \left(\sum_{\alpha \in \mathcal{N}
     } \alpha>3\right)\right),\label{r2}\\
     f(0,\{\alpha\})&=&1 \textrm{ if } \left( \sum_{\alpha \in \mathcal{N}
      } \alpha= 3\right),\label{r3}\\
      f(0,\{\alpha\})&=&0 \textrm{ if } \left (\sum_{\alpha\in \mathcal{N}
       } \alpha\neq3\right).\label{r4}
 \end{eqnarray}
Here   $\{\alpha\}$ is the set of all state values in $\mathcal{N}$, and $|\mathcal{N}|$ is the cardinality of $\mathcal{N}$. These rules mimic the basic processes of life and death: rule \eqref{r1} represents
 condition for sustainable life, rule \eqref{r2} represents death due to under or over population, rule \eqref{r3} represents a birth condition,
 and rule \eqref{r4} corresponds to the permanence of a death state condition. Cells are generally updated synchronously, \emph{i.e.},  they undergo state transitions at the same time, although in some papers there are  variations implementing also asynchronous evolutions (see, for instance, \cite{LAPM}).

{
The use of quantum ideas for cellular automata (QCA) dates back to the 1980's  (\cite{Fey,Deu,GZ}), and has attracted the interest of several scientists during the last decades. The motivation behind these  approaches mainly relies on the possibility of reproducing, by using generalized structures, quantum phenomena such as interference, or entanglement effects. In this context, various \emph{quantum} versions of the game of life have been developed. In \cite{FA}, by using standard arguments in the QCA, the state $|\psi\rangle$ of a cell is defined as a superposition of the states $|1\rangle$ (life) and $|0\rangle$ (death), forming a qubit $|\psi\rangle=c_1|1\rangle +c_0|0\rangle$, and the process of birth-death-sustain of a cell is  reproduced through the combination of suitable birth and death operators. A different quantum version of the game of life has been analyzed in \cite{AG} in the context of the so-called {\em universal} and {\em partitioned} QCA. Still another approach, based on the number operator, and involving an Hamiltonian operator which includes mechanisms resembling the standard rules of the game of life, is developed in \cite{BCM2012}.

The quantum version of the \emph{GoL} introduced in this paper, hereafter \emph{QGoL}, is not intended as an attempt to study any quantum property of the QCA, but just as a proposal of a deterministic method describing the structure of peculiar cellular automata by means of an enriched concept of \emph{rule}. In particular, we suppose that, during consecutive transients, the system is driven by an energy-like operator, describing the most relevant mechanisms occurring in the system itself.
The main idea behind this approach is based on methods typically connected with
quantum mechanics, but recently adopted also for the analysis  of several macroscopic systems. We just mention some application in  social life and decision
making processes \cite{qdm1,qdm2}, in population dynamics  \cite{ff2,fff}, and in ecological processes \cite{BCO16, DO16}.
More in details, since according to \cite{bagbook} the dynamical variables representing the whole system are assumed to be operator-valued, the dynamics is deduced by introducing the self-adjoint operator $H$ (the energy of the
system) containing the effects of all possible interactions between the different
parts of the physical system.} Therefore, differently from the \emph{GoL}, we shall consider a quantic dynamics of the population before applying the rule
$\rho$. Moreover, $\rho$ somehow extends the rule introduced in (\ref{r1})-(\ref{r4}), and, in fact, may be rather general. The new state deduced after the rule is implemented is then considered as the starting point for the next iteration of the time evolution, which is again driven by $H$. At the end of this new iteration, $\rho$ is applied once more, and a new state is deduced. And so on.  Of course, the dynamics one deduces in this way is driven by several ingredients, and, in particular, by the Hamiltonian $H$, by the rule $\rho$, and by the initial status of the system $\Sc$. We shall refer to the whole procedure as the $(H,\rho)$--{\em induced dynamics} of $\Sc$. Our first interest here is to produce a general mathematical setting in which this \emph{QGoL} can be well discussed, and then to apply this procedure to a concrete situation and describe the possible scenarios that can arise.

The paper is organized as follows. In  Section~\ref{sect2}, we describe the general mathematical framework for an
$(H,\rho)$--induced dynamics. In Section~\ref{sect3}, we describe the dynamics of  a \emph{QGoL} ruled by  a strictly quadratic Hamiltonian.  This choice is technically useful, since it produces linear differential equations which can be explicitly solved
(see \cite{ff2,fff,bagbook}). In Section~\ref{sec:results},  we analyze in detail our results by means of different statistical tools; in particular, we perform a spectral analysis to study the influence of the various parameters in the $(H,\rho)$--induced dynamics and the differences with respect to the classical \emph{GoL}; successively, we consider a {\em blob analysis} of the model, looking again for differences and similarities between \emph{QGoL} and \emph{GoL}. Our conclusions are given in Section~\ref{sect5}. In the Appendix, we  present a detailed analysis on the formation of periodic solutions of the problem introduced in Section~\ref{sect3} in a small domain.

\section{The general setting}\label{sect2}

In this Section, we introduce, at a rather general level, our idea of $(H,\rho)$--induced dynamics. As it will appear clear from our treatment, this idea merges the general framework of  quantum dynamics with the possibility that the dynamics may be periodically disturbed because of some external (or internal) action, whose effects are not easily described by any self-adjoint Hamiltonian operator.

Let $\Sc$ be our physical system and $Q_j$ ($j=1,\ldots,M$) a set of $M$ commuting self-adjoint operators with eigenvectors $\varphi^{(j)}_{\alpha_n}$ and eigenvalues $\alpha_n^{(j)}$:
\be
[Q_j,Q_k]=0, \qquad Q_j=Q_j^\dagger,\qquad  Q_j\varphi^{(j)}_{n_j}=\alpha_{n_j}^{(j)}\varphi^{(j)}_{n_{j}},
\label{21}
\en
$j,k=1,2,\ldots,M$, $n_j=1,2,3,\ldots,N_j$, which can be finite or infinite.
We set $\mathbf{n}=(n_1,n_2,\ldots,n_M)$, and
$$
\varphi_{\mathbf{n}}=\varphi^{(1)}_{n_{1}}\otimes\varphi^{(2)}_{n_{2}}\otimes\cdots\varphi^{(M)}_{n_{M}}.
$$
This is an eigenstate of all the operators $Q_j$:
\be
Q_j\,\varphi_{\mathbf{n}}=\alpha_{n_j}^{(j)}\,\varphi_{\mathbf{n}}.
\label{22}
\en
The existence of a common eigenstate for all the operators $Q_j$ is guaranteed by the fact that they mutually commute. It is convenient, and always true in our applications, to assume that these vectors are mutually orthogonal and normalized:
\be
\left<\varphi_{\mathbf{n}},\varphi_{\mathbf{m}}\right>=\delta_{\mathbf{n},\mathbf{m}}=\prod_{j=1}^M\delta_{n_j,m_j}.
\label{23}
\en
The Hilbert space $\Hil$ where $\Sc$ is  defined is (mathematically) constructed as the closure of the linear span of all the vectors $\varphi_\mathbf{n}$, which therefore turn out to form an orthonormal basis for $\Hil$. Now, let $H=H^\dagger$ be the time-independent self-adjoint Hamiltonian of $\Sc$, which, in general, does not commute with the $Q_j$'s. This means that, in absence of any other information, the wave function $\Psi(t)$ describing $\Sc$ at time $t$ evolves according to the
Schr\"odinger equation $i\dot\Psi(t)=H\Psi(t)$, where $\Psi(0)=\Psi_0$ describes the initial status of $\Sc$. It is well known \cite{mer,mes} that this is not the unique way to look at the time evolution of $\Sc$. Another equivalent way consists in adopting the Heisenberg representation, in which the wave function does not evolve in time, while the operators do, according to the Heisenberg equation $ \dot X(t)=i[H,X(t)]$. Here $X(t)$ is a generic operator acting on $\Hil$, at time $t$, and $[A,B]=AB-BA$ is the commutator between $A$ and $B$. In this paper, we will mostly adopt the first point of view, \emph{i.e.}, we use essentially the Schr\"odinger representation.
The formal solution\footnote{The reason why we speak about a formal solution is that $\exp(-iHt)\Psi_0$ is not, in general, explicitly known, at least if there is no easy way to compute the action of the unitary operator $\exp(-iHt)$ on the vector $\Psi_0$, which is not granted at all. This is not very different from the equivalence of a differential equation with some given initial conditions and its integral counterpart: they contain the same information but none of them provide the explicit solution of the dynamical problem.} of the Schr\"odinger equation is, since $H$ does not depend explicitly on $t$, $\Psi(t)=\exp(-iHt)\Psi(0)=\exp(-iHt)\Psi_0$.
We can now compute the mean value of each operator $Q_j$ in the state $\Psi(t)$: $q_j(t)=\left<\Psi(t),Q_j\Psi(t)\right>$, and use it to define the related $M$-dimensional time-dependent vector $\mathbf{q}(t)=(q_1(t),q_2(t),\ldots,q_M(t))$.

We are now ready to introduce, rather generally, the notion of {\em rule} $\rho$ as a map from $\Hil$ to $\Hil$.
This rule is not necessarily linear, and its explicit action depends on the expression of $\mathbf{q}(t)$ at particular instants $k\tau$ ($k=1,2,\ldots$). In other words, according to how $\mathbf{q}(k\tau)$ looks like, $\rho$ maps an input vector $\Phi_{in}$ into a different output vector $\Phi_{out}$, and we write  $\rho(\Phi_{in})=\Phi_{out}$\footnote{Maybe, a more precise notation should be $\rho_{\mathbf{q}(k\tau)}(\Phi_{in,k})=\Phi_{out,k+1}$, but we prefer to use the above notation.}.
This is not very different from what happens in scattering theory, where an incoming state, after the occurrence of the scattering, is transformed into an outgoing state \cite{roman}.

\subsection{The rule $\rho$ in the induced dynamics}\label{subsec:rulerho}
The rule, up to this moment, has been introduced in a very general way as a map from $\Hil$ to $\Hil$;  nevertheless, in view of our concrete application in Section~\ref{sect3}, we now discuss a special definition of the rule which is suitable for our purposes. At first, we observe that there exists a one-to-one correspondence between $\mathbf{n}$ and the vector $\varphi_\mathbf{n}$:
once we know $\mathbf{n}$, $\varphi_\mathbf{n}$ is clearly identified, and  viceversa. Suppose now that at
time $t=0$ the system $\Sc$ is in a state $\mathbf{n}^0$ or, which is the same, $\Sc$ is described by the vector
$\varphi_{\mathbf{n}^0}$. Then, once fixed a positive value of $\tau$, this vector evolves
 in the time interval $[0,\tau[$ according to the Schr\"odinger recipe: $\exp(-iHt)\varphi_{\mathbf{n}^0}$. Let us set
\[
 \Psi(\tau^-)=\lim_{t\rightarrow \tau^-}\exp(-iHt)\varphi_{\mathbf{n}^0},
\]
where $t$ converges to $\tau$ from below\footnote{We use here $\tau^-$, $2\tau^-$, $\ldots$,  as argument of $\Psi$ to emphasize that {\bf before} $\tau^-$, for instance, the time evolution is only due to $H$, while $\rho$ really acts at $t=\tau$.}. Now, at time $t=\tau$, $\rho$ is applied to $\Psi(\tau^-)$, and the output of this action is a
new vector which we assume here to be again an eigenstate of each operator $Q_j$, but with different eigenvalues, $\varphi_{\mathbf{n}^1}$\footnote{This choice is not the only possibility to set up a rule. In fact, other possibilities can also be considered. The key common point to all possible choices is that $\rho$ behaves as a check over the system $\Sc$, and modifies some of its ingredients according to the result of this check.}. In other words, $\rho$ {\em looks } at the  explicit expression of the vector $\Psi(\tau^-)$ and, according to its form, returns a new vector $\mathbf{n}^1=(n^1_1,n^1_2,\ldots,n^1_{M})$;  as a consequence, a
new vector $\varphi_{\mathbf{n}^1}$ of $\Hil$ is obtained. Examples of how $\rho$ explicitly acts are given in Sections \ref{sectAfirstappl} and \ref{sect3}. Now, the procedure is iterated, taking $\varphi_{\mathbf{n}^1}$
as the initial vector, and letting it evolve with $H$ for another time interval of length $\tau$;
we compute
$$
\Psi(2\tau^-)=\lim_{t\rightarrow \tau^-}\exp(-iHt)\varphi_{\mathbf{n}^1},
$$
and the new vector  $\varphi_{\mathbf{n}^2}$ is deduced by the action of rule $\rho$ on $\Psi(2\tau^-)$: $\varphi_{\mathbf{n}^2}=\rho(\Psi(2\tau^-))$.   Then, in general, we have
\be
\Psi(k\tau^-)=\lim_{t\rightarrow \tau^-}\exp(-iHt)\varphi_{\mathbf{n}^{k-1}},
\label{add1}
\en
and
\be
\varphi_{\mathbf{n}^k}=\rho \left(\Psi(k\tau^-)\right),
\label{add2}
\en
for all $k\geq1$.

Let now $X$ be a generic operator on $\Hil$, bounded or unbounded. In this last case, we will require that the various $\varphi_{\mathbf{n}^k}$  belong to the domain of $X(t)=\exp(iHt)X\exp(-iHt)$ for all $t\in[0,\tau]$. For later convenience, it is useful to observe that this condition is satisfied in the \emph{QGoL}.

\begin{defn}\label{def1}
The sequence of functions
\be
x_{k+1}(t):=\left<\varphi_{\mathbf{n}^k}, X(t)\varphi_{\mathbf{n}^k}\right>,
\label{24}\en
for $t\in[0,\tau]$ and $k\in {\Bbb N}_0$, is called the $(H,\rho)$--induced dynamics of $X$.
\end{defn}
It is clear that $x_{k+1}(t)$ is well defined, because of our assumption on $\varphi_{\mathbf{n}^k}$. In particular, suppose that $X$ is a bounded positive operator. Then $X$ can be written as $X=A^\dagger A$, for a suitable bounded operator $A$ \cite{rs}. Hence, it is easy to check that each $x_{k+1}(t)$ is non-negative for all allowed $t$ and $k$:
\be
\begin{aligned}
x_{k+1}(t)&=\left<\varphi_{\mathbf{n}^k}, \exp(iHt)(A^\dagger A)\exp(-iHt)\varphi_{\mathbf{n}^k}\right>=\\
&=\left\|A\exp(-iHt)\varphi_{\mathbf{n}^k}\right\|^2\geq0.
\end{aligned}
\label{evolutionN}
\en

Some properties of the sequence ${\underline X}(\tau)=(x_1(\tau),x_2(\tau),x_3(\tau),\ldots)$, arising from the $(H,\rho)$-induced dynamics of a given operator $X$, can be easily proved.

\begin{prop}\label{prop1}
The following  results concerning periodicity hold true.

\begin{enumerate}

\item If the rule $\rho$ does not depend on the input, then $${\underline X}(\tau)=\left(x_1(\tau),x_2(\tau),x_2(\tau),x_2(\tau),x_2(\tau),\ldots\right).$$

\item Assume that a $K>0$ exists such that $\rho(\varphi_{\mathbf{n}^K})=\varphi_{\mathbf{n}^K}$, then
$${\underline X}(\tau)=\left(x_1(\tau),x_2(\tau),\ldots,x_{K+1}(\tau),x_{K+1}(\tau),x_{K+1}(\tau),\ldots\right).$$

\item  Assume that  $K>0$, $N\geq0$ exist such that $\rho(\varphi_{\mathbf{n}^{(N+K)}})=\varphi_{\mathbf{n}^N}$, then
$${\underline X}(\tau)=\left(x_1(\tau),\ldots,x_N(\tau),x_{N+1}(\tau),\ldots,x_{N+K+1}(\tau),x_{N+1}(\tau),\ldots\right).$$

\end{enumerate}
\end{prop}

The proofs of all these statements are easy consequences of the definition of $(H,\rho)$--induced dynamics, and of how the rule works. It is clear that more situations of this kind can still be deduced, other than the ones given by the Proposition above, but we will not discuss  them here. On the other hand, we want to notice that from ${\underline X}(t)=(x_1(t),x_2(t),x_3(t),\ldots)$ it is possible to define a function of time in the following way:
\be
\tilde X(t)=
\left\{
\begin{array}{ll}
x_1(t),\qquad\qquad t\in [0,\tau[ &  \\
x_2(t-\tau),\qquad t\in [\tau,2\tau[ &  \\
x_3(t-2\tau),\qquad t\in [2\tau,3\tau[ &  \\
\ldots &
\end{array}%
\right.
\label{25bis}
\en
It is clear that $\tilde X(t)$ may have discontinuities in $k\tau$, for positive
integers $k$.  Of course, Proposition \ref{prop1} gives conditions for $\tilde X(t)$ to admit some asymptotic value or to be periodic. We will consider this aspect later on.

\vspace{2mm}

Let us now discuss the operator representation of $\rho$. As we will show later, this representation  produces a bounded operator. Let $f,g,h\in\Hil$ be three vectors of the Hilbert space $\Hil$. We set $f\otimes\overline{g}\,(h):=\left<g,h\right>\,f$. So, $f\otimes\overline{g}$ projects any vector along $f$. Then we introduce the operator $R$ as follows:
\be
R=\sum_{k\geq0} \varphi_{\mathbf{n}^{k+1}}\otimes \overline{\exp(-iH\tau)\varphi_{\mathbf{n}^k}}.
\label{26}
\en

The operator is a finite sum of simple rank one operators, and the number of its addenda clearly depends on the time interval we are interested to. So $\|R\|\leq N$, where $N$ are the number of contributions in the sum in (\ref{26}). It is clear that $R\left(\exp(-iH\tau)\varphi_{\mathbf{n}^k}\right)=\varphi_{\mathbf{n}^{k+1}}$.
This is a consequence of the fact that $\exp(-iH\tau)$ is unitary and that the various $\varphi_{\mathbf{n}^k}$ are mutually orthogonal.  Then it is clear that $R\left(\exp(-iH\tau)\varphi_{\mathbf{n}^k}\right)=\rho\left(\exp(-iH\tau)\varphi_{\mathbf{n}^k}\right)$, while it is not granted \emph{a priori} that $\rho(\Phi)=R\Phi$ for a generic vector $\Phi$ in $\Hil$. For this reason, $R$ can only be thought as an {\em effective} representation of $\rho$.

The operator $R$ can be slightly simplified if some of the assumptions of Proposition \ref{prop1} apply. For instance, if for some $j$ we have $\mathbf{n}^j=\mathbf{n}^{j+1}=\mathbf{n}^{j+2}=\cdots$, then
$$
R=\sum_{k=0}^{j-1} \varphi_{\mathbf{n}^{k+1}}\otimes \overline{\exp(-iH\tau)\varphi_{\mathbf{n}^k}}+(N-j)\varphi_{\mathbf{n}^{j}}\otimes \overline{\exp(-iH\tau)\varphi_{\mathbf{n}^j}}
$$
which in particular, if $j=1$, becomes quite simple:
\be
R=\varphi_{\mathbf{n}^{1}}\otimes \overline{\Phi},
\label{26bis}\en
where $\Phi=\exp(-iH\tau)\left(\varphi_{\mathbf{n}^{0}}+(N-1)\varphi_{\mathbf{n}^{1}}\right)$.

An obvious remark about $R$ is that it can only be found \emph{a posteriori}. In fact, because of its definition (\ref{26}), $R$ is known when the various $\varphi_{\mathbf{n}^k}$ are known, but these can only be deduced using (several times) $\rho$. Hence, in order to write $R$, we have to use $\rho$. In other words, equation (\ref{26}) is not really useful to deduce, for instance, the time evolution of $\Sc$. What is true is exactly the opposite:
\emph{it is the constrained time evolution of $\Sc$ which determines the expression for $R$}.

\begin{remark} If we look at the standard  \emph{GoL} the time plays no role: what is really relevant is the rule $\rho$. This can be easily recovered, in our scheme, just taking $\tau=0$, or assuming that $H=0$. In both cases the sequence of functions defined above produces a sequence of (in general) complex numbers ${\underline X}=(x_1,x_2,x_3,\ldots)$, where $x_j=x_j(0)= \left<\varphi_{\mathbf{n}^{j-1}}, X\varphi_{\mathbf{n}^{j-1}}\right>$, $j\geq1$. Hence, our strategy contains two limiting cases: if $H=0$ or $\tau=0$ then we recover the standard \emph{GoL}, as commonly discussed in the literature. On the other hand, if we assume that $\Phi_{out}=\rho(\Phi_{in})=\Phi_{in}$ for all $\Phi_{in}$, we are essentially saying we have no rule at all, and we go back to the standard quantum dynamics.
\end{remark}

\subsection{A first application}\label{sectAfirstappl}

\begin{figure}
\begin{center}
\includegraphics[width=0.7\textwidth]{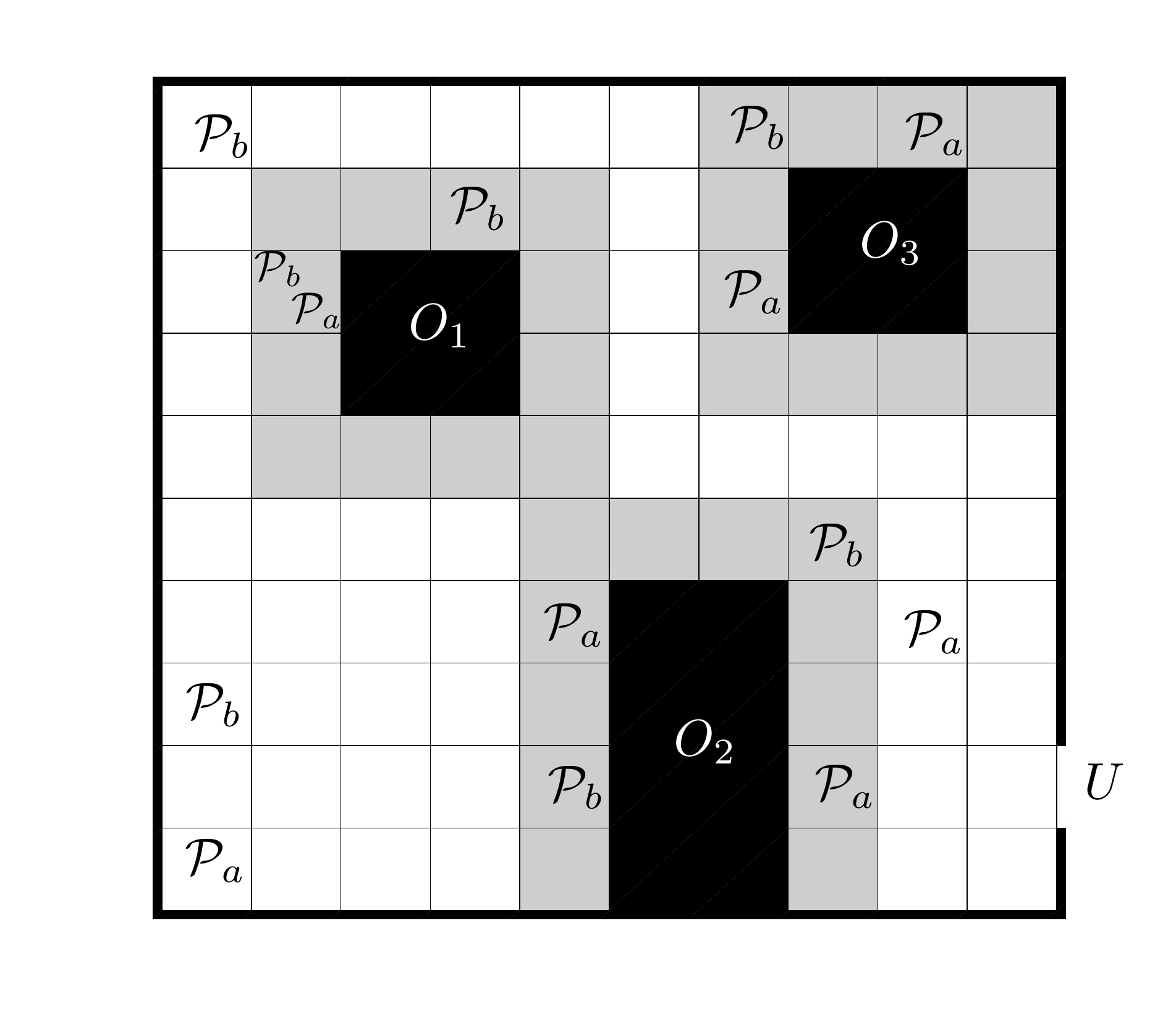}
\end{center}
\caption{\label{setup_p2_u1_po}
\footnotesize The room $\R$ is a square of $L_x\cdot L_y=10\cdot10$ cells with an exit cell $U$ and three obstacles $O_1$, $O_2$, $O_3$, surrounded by a region
$\partial O$, which might be slightly different from the rest of $\R$. At $t=0$ the populations $\Pc_a$ and $\Pc_b$ are distributed as shown in figure.
}
\end{figure}

In a recent paper \cite{fff}, the general scheme discussed so far was applied to the analysis of a particular problem in the dynamics of crowds. The aim of that paper was to propose an analysis of the escape strategies of a number of people originally localized in a room with some obstacles and some exits. The goal was to minimize the time needed by the people to leave the room. This is clearly  of a certain interest in the case of some alarm. Here, we just want to sketch some aspects of that model, and in which sense it is close to what we propose here.
Assume we have two populations, $\Pc_a$ and $\Pc_b$, inside a room $\R$ with a single exit $U$ and some obstacles $O_j$, as in Figure \ref{setup_p2_u1_po};  the distributions of  $\Pc_a$ and $\Pc_b$ are also shown. As we can see from the figure, both $\Pc_a$ and $\Pc_b$ occupy, at $t=0$, seven (mostly) different cells of $\R$.

We refer to \cite{fff} for the analytic form of the Hamiltonian $H$ which describes the dynamics of the two populations without any rule. $H$ includes a free dynamics, an interaction between $\Pc_a$ and $\Pc_b$, and the diffusion of these populations along $\R$. In \cite{fff}, the time evolution of the system is given in the Heisenberg, rather than in the Schr\"odinger, representation. The reason is that this was a natural and efficient way to describe the time evolution of the density of the populations in each part of the room. However, see \cite{mer,mes}, it is well known that the two representations are unitarily equivalent, and using one rather than the other is only a matter of convenience. Since the total densities of the populations inside $\R$ commute with $H$, the system (if we do not add any rule) is not suitable to describe people leaving the room. On the contrary, this model would describe a movement of $\Pc_a$ and $\Pc_b$ {\bf inside} the room. A mechanism was then introduced to break down this densities conservation: after a fixed time interval (corresponding to our time $\tau$), a check on the densities of the populations reaching the cell $U$ (the exit!) is performed. If these densities are below a certain threshold, then nothing changes in the system, otherwise, after the check, the originally (\emph{i.e.},
at $\tau^-$) high density, is set equal to zero at time $\tau$. This is an efficient way to describe the fact that, when the people have reached the exit $U$, they do not enter again the room! They just want to \emph{disappear}. And this is exactly what our
rule $\rho$ does here. We will be more explicit on the definition of the rule in
Section~\ref{sect3}, where the GoL is discussed in details.

\subsection{On equilibria}

In view of the applications to the \emph{QGoL}, it is useful to introduce now the following definitions, which are directly connected with Proposition~\ref{prop1}.

\begin{defn}\label{def2}\
\begin{enumerate}
\item  $x^\infty\in \Bbb C$ is an equilibrium for the $(H,\rho)$--induced dynamics of the operator $X$ if, $\forall\,\epsilon>0$, $\exists N_\epsilon>0$ such that $\left|x_\ell(\tau)-x^\infty\right|<\epsilon$, for all $\ell>N_\epsilon$.

\item Given $\epsilon>0$, $x^\infty\in \Bbb C$ is an $\epsilon$-equilibrium for the $(H,\rho)$-induced dynamics of the operator $X$ if  $\exists N_\epsilon>0$ such that $\left|x_\ell(\tau)-x^\infty\right|<\epsilon$, for all $\ell>N_\epsilon$.

\item The $L$-dimensional vector $\left(x_1^\infty,x_2^\infty,\ldots, x_L^\infty\right)$, $x_j^\infty\in \Bbb C$, is an $L$-equilibrium cycle for the $(H,\rho)$-induced dynamics of the operator $X$ if, $\forall\,\epsilon>0$, $\exists N_\epsilon>0$ such that $$\sup_{\ell=1,2,\ldots,L}\left|x_{N_\epsilon+kL+\ell}(\tau)-x_\ell^\infty\right|<\epsilon,$$ for all $k=0,1,2,\ldots$.\\  In this case we call $L$ the period of the solution and $N_L=inf_{\epsilon>0} N_{\epsilon}$ the transient  to reach the $L$-equilibrium cycle.

\end{enumerate}
\end{defn}

\begin{remark}
If $x^\infty$ is an equilibrium for the $(H,\rho)$-induced dynamics of the operator $X$, then it is also an $\epsilon$-equilibrium, for all $\epsilon>0$.
\end{remark}
\begin{remark}
According to the definition given in Proposition~\ref{prop1}, a $1$-equilibrium cycle solution for the $(H,\rho)$-induced dynamics of the operator $X$ is simply an equilibrium. We could also extend the definition of $\epsilon$-equilibrium to cycles, but this is not interesting for us and will not be done here.
\end{remark}

The following results easily follow from Definition~\ref{def2} and
Proposition~\ref{prop1}.

\begin{prop} If the rule $\rho$ does not depend on the input, or, more in general, if there exists $K>0$ such that $\rho(\varphi_{\mathbf{n}^K})=\varphi_{\mathbf{n}^K}$, then, for each operator $X$ of the system, an equilibrium for the $(H,\rho)$-induced dynamics of the operator $X$ does exist.

Suppose rather that a $K>0$ exists such that $\rho(\varphi_{\mathbf{n}^K})=\varphi_{\mathbf{n}^0}$, and let us define $\overline{\epsilon}=\max_{j=1,2,\ldots,K}\left|x^\infty-x^j(\tau)\right|$, with $x^\infty=\frac{1}{K}\,\sum_{j=1}^Kx^j(\tau)$, then $x^\infty$ is an $\overline{\epsilon}$-equilibrium for the $(H,\rho)$-induced dynamics of the operator $X$.

\end{prop}

Once again, we do not give here the proof of the Proposition, which is very easy. It is clear that the interesting situation is when $\overline{\epsilon}$ is sufficiently small. When this is not so, we can not say much about the closeness of $x^\infty$ to the various $x^j(\tau)$. In this case, it is more interesting the following result.

\begin{prop} Suppose that $K>0$ exists such that $\rho(\varphi_{\mathbf{n}^K})=\varphi_{\mathbf{n}^0}$, then a $(K+1)$-cycle for the $(H,\rho)$-induced dynamics of the operator $X$ exists, with $x_j^\infty=x^j(\tau)$.

\end{prop}

It is clear that, even if an equilibrium exists for the $(H,\rho)$-induced dynamics of a certain operator $X$, then not necessarily it is  an equilibrium also for the $(H,\rho)$-induced dynamics of a different operator $Y$. In other words, using the function $\tilde X(t)$ introduced before, even if this function can admit some asymptotic value (or being periodic from some multiple of $\tau$), the analogous function $\tilde Y(t)$ defined in analogy to (\ref{25bis}) does not necessarily admit some asymptotic value.

Also, it is easy to understand that, in presence of some equilibrium, the operator $R$ in (\ref{26}) admits a simpler form, as for instance that in (\ref{26bis}), for a suitable vector $\Phi$.

\section{The quantum game of life}\label{sect3}

In this Section, we introduce a variant of the classical \emph{GoL} by using at each new generation the $(H,\rho)$--induced quantum dynamics described in the previous Section. In particular, to each cell of the lattice is attached  a fermionic variable, taking value 0 or 1 only\footnote{Equivalently, we could use spin variables and work with Pauli matrices.}, and each possible configuration is given as a vector on the Hilbert space $\Hil$ described below. The quantum approach we want to describe is based on the assumption that the \emph{observables} of the system $\Sc$ we are interested to, among which there is the state of each cell, are described by operators acting on  $\Hil$.

We suppose that the system $\Sc$ is made by a single population $\Pc$ living on a square lattice $\R$ made by $L^2$ cells. At time zero, a cell may be dead or alive (these states are represented by the values 0 or 1, respectively). This setting is well described by using a two state vector $\varphi_{n_\alpha}$ to describe the cell, where $\alpha$ labels the cell, and $n_\alpha=0,1$. A simple way to build up these vectors (one for each cell) is to introduce a family of fermionic operators, one for each $\alpha$, \emph{i.e.}, a family of operators $a_\alpha$ satisfying the following canonical anticommutation rules (CAR):
$$
\{a_\alpha,a_\alpha^\dagger\}=a_\alpha a_\alpha^\dagger+a_\alpha^\dagger a_\alpha=\1, \qquad a_\alpha^2= (a_\alpha^{\dagger})^2=0.
$$
The operators $a_\alpha, a_\alpha^\dagger$  are  the fermionic  \emph{annihilation} and \emph{creation} operators,  respectively. These operators are very well known and widely
analyzed in any textbook
on quantum mechanics (see, for instance, \cite{roman}); hence, here we only briefly recall some of their properties useful for our purposes.

From $a_\alpha, a_\alpha^\dagger$ we can construct  the operator $N_\alpha=a_\alpha^\dagger a_\alpha$, which is the number operator for the cell $C_\alpha$, and $\varphi_{0_\alpha}$, which is the vacuum of $a_\alpha$, \emph{i.e.}, the vector satisfying $a_\alpha\varphi_{0_\alpha}=0$. Moreover, $\varphi_{1_\alpha}$ is simply $a_\alpha^\dagger\varphi_{0_\alpha}$, and $N_\alpha\varphi_{n_\alpha}=n_\alpha\varphi_{n_\alpha}$, with $n_\alpha=0,1$. In this way, we have exactly the two vectors we were looking for, and the eigenvalues of the operator $N_\alpha$ describing the status of the cell (dead or alive).
Then, we define the state vector of the system as
 \bea
 \varphi_\mathbf{n}=\otimes_{\alpha=1}^{L^2}\varphi_{n_\alpha}, \quad \mathbf{n}=(n_1,n_2,...,n_{L^2}),\label{vectorstate}
 \ena
which clearly describes the status of each cell in $\R$.
The Hilbert space $\Hil$ is constructed by taking the closure of the linear span of all these vectors. The scalar product is the natural one. In particular, in each cell the scalar product reduces to the one in ${\Bbb C}^2$. The CAR in $\R$ extend those above:
\be
\{a_\alpha,a_\beta^\dagger\}=\delta_{\alpha,\beta}\1, \qquad \forall \alpha,\beta,
\label{31}\en
where $a_\alpha$ is now a $2^ {L^2}\times2^{L^2}$ matrix operator satisfying
\beano
a_\alpha \varphi_{\mathbf{n}}&=&0 \quad \textrm{if } n_{\alpha=0},\\
a_\alpha^\dagger \varphi_{\mathbf{n}}&=&0 \quad \textrm{if } n_{\alpha=1},\\
N_{\alpha}\varphi_{\mathbf{n}}&=&a_\alpha^\dagger a_\alpha\varphi_{\mathbf{n}}=n_{\alpha}\varphi_{\mathbf{n}}.
\enano
The general Hamiltonian describing the diffusion of a population in a closed region through fermionic operators (see \cite{fff,gar}) is assumed here to be
\be
H= \sum_{\alpha=1}^{L^2} a^{\dagger}_{\alpha}a_{\alpha}+\sum_{\alpha,\beta=1}^{L^2}  p_{\alpha,\beta}\left(a_{\alpha} a_{\beta}^\dagger+a_{\beta} a_{\alpha}^\dagger\right),
\label{hamiltonian1Pop}
\en
where  $p_{\alpha,\beta}$ are non-negative real parameters such that
$p_{\alpha,\beta}=1$ if $\alpha\neq\beta$ are neighboring cells\footnote{We consider for each cell the Moore neighborhood made, for internal cells, of the eight surrounding cells. Less cells obviously form the Moore neighborhood of a cell on the border.},  and $p_{\alpha,\beta}=0$  otherwise.
Note that $H$ is self-adjoint, \emph{i.e.}, $H=H^{\dag}$.  We notice that in \cite{fff} the parameters $p_{\alpha,\beta}$ could take any positive real value, and not only zero and one. In this way, the speed of  diffusion from one cell to another could be changed. However, to simplify our discussion, we avoid this possibility here.

We introduce now the essential variation with respect to the classical \emph{GoL}: in fact, before the
generation of a new state, we fix a {\em transient time} $\tau$ such that in the time interval $[0,\tau[$
the neighboring cells interact  in a way which is driven by the Hamiltonian
$H$ given in (\ref{hamiltonian1Pop}); hence, as time $t$ increases, $t<\tau$, $\Sc$ is no more in its initial
state, $\phi_{\mathbf{n}^0}$, but instead in the evolved state $\exp(-iHt)\phi_{\mathbf{n}^0}$ { which, in general, is a superposition of the vectors $\varphi_{\bf n}$ defined in \eqref{vectorstate}}. Following the scheme
described in \cite{bagbook}, we relate the mean values of the number operators $N_{\alpha}$ to the new
states of each cell. Using \eqref{24}-\eqref{evolutionN}, where the Hamiltonian $H$ is given in
\eqref{hamiltonian1Pop}, we recover the  evolution of the number operators  as
$$N_\alpha(t):=\exp(iHt)N_\alpha(0) \exp(-iHt),$$
and then their mean values on some suitable state $\phi_{\mathbf{n}^0}$ describing the system at $t=0$, as
\begin{equation}
n_{\alpha,0}(t)=\langle\phi_{\mathbf{n}^0},N_\alpha(t) \phi_{\mathbf{n}^0} \rangle=\lVert  a_\alpha \exp(-iHt)(t)
\varphi_{\mathbf{n}^0} \rVert^2.
\label{evolutionN0}
\end{equation}

Because of the CAR, the values $n_{\alpha,0}(t)$ belong to the range $[0,1]$, for all $\alpha$ and all $t$.
Hence, they can be endowed with a probabilistic meaning: for instance, if $n_{\alpha,0}(t) \ll 1$ then  the cell
$\alpha$ has high probability to be in a dead state. We let $t$ vary in the interval $[0,\tau[$. Then,  at time
$\tau$, we apply the rules synchronously to all the cells, so that the upgraded states are all either 0 or 1, and
the  new  state vector, obtained through \eqref{vectorstate}, is  $\varphi_{\mathbf{n}^1}$. This process is
iterated for several generations. The whole  procedure can be schematized as follows.

\begin{algorithmic}
\LOOP[From generation $k$ to generation $k+1$]
\STATE $\bullet$ For each cell $\alpha$ set $n_{\alpha,k}(0)=n_{\alpha,k}$, with $n_{\alpha,k}=0\textrm{ or }
1$, and construct $\varphi_{\mathbf{n}^k}$ through \eqref{vectorstate}.
\STATE $\bullet$ Compute $\exp(-iH\tau)\varphi_{\mathbf{n}^k}$ and the related $n_{\alpha,k}(\tau)$, in analogy with (\ref{evolutionN0}), out of it.
\STATE $\bullet$ Apply the rule synchronously to $\exp(-iH\tau)\varphi_{\mathbf{n}^k}$ to compute $\mathbf{n}^{k+1}$. In  each cell we will have
$n_{\alpha,k+1}=0\textrm{ or }1$.
\STATE $\bullet$ Set $k\rightarrow k+1$.
\ENDLOOP
\end{algorithmic}

Hence, we obtain in each cell a sequence of states $n_{\alpha,k}$, where the index $k$ labels the generic
$k$-th generation. The way in which, at each generation, $n_{\alpha,k}$ is set to $0$ or $1$ is governed by the
rules we want to apply, which are an extended version of the ones described in Section~\ref{sec:intro}. More
explicitly, our  rules $\rho$  for the generation of the new state are defined as follows:
\begin{eqnarray}
&&\rho_{\sigma}(n_{\alpha,k}=1)=1 \textrm{ if } \left(2-\sigma\leq \sum_{\beta \in \mathcal{N}
 } n_{\beta,k}\leq 3+\sigma\right),\label{nr1}\\
 &&\rho_{\sigma}(n_{\alpha,k}=1)=0 \textrm{ if } \left (\sum_{\beta \in \mathcal{N}
  } n_{\beta,k}<2-\sigma\right) \vee  \left(\sum_{\beta \in \mathcal{N}
    } n_{\beta,k}>3+\sigma\right),\label{nr2}\\
&&\rho_{\sigma}(n_{\alpha,k}=0)=1 \textrm{ if } \left(3-\sigma\leq \sum_{\beta \in \mathcal{N}
     } n_{\beta,k}\leq 3+\sigma\right),\label{nr3}\\
&&\rho_{\sigma}(n_{\alpha,k}=0)=0 \textrm{ if }\left (\sum_{\beta \in \mathcal{N}
       } n_{\beta,k}<3-\sigma\right) \vee  \left(\sum_{\beta \in \mathcal{N}
         } n_{\beta,k}>3+\sigma\right),\label{nr4}
         \label{quantum_rules}
\end{eqnarray}
where $\sigma$ is a positive parameter, which can be seen as a measure of
the deviation from the original
classical rule. In particular, if $\sigma=0$, we recover essentially the rule given by (\ref{r1})-(\ref{r4}).
Through this procedure we obtain a sequence of functions $n_{\alpha,k}(t)$ with $t\in[0,\tau[$ which define the
$(H,\rho)$-induced dynamics for the various number operators $N_{\alpha}$ as in Definition \ref{def1}.

\vspace{2mm}

{\bf Remark:--}
It is worth mentioning that the quantum version of the game of life proposed in \cite{BCM2012} has some similarities with our approach, in particular for the use of a suitable Hamiltonian operator and of number operators to count the densities of the cells. Furthermore, these densities are computed through the expectation values of the number operators (as we do too), and  a statistical comparison  with the classical game of life in the 1D-case is performed in terms of global mean density  and \textit{diversity} of the cells. In a very schematic way, using our notation, they consider the evolution of an initial state expressed by \eqref{vectorstate} driven by the following Hamiltonian:
\beano
&&H_{BCM}=\sum_\alpha (a^{\dagger}_{\alpha}+a_{\alpha})(h_{\alpha}^l+h_{\alpha}^d),\\
&&h_{\alpha}^l=\sum_{\mathcal{N_\alpha}}N_{q_1}N_{q_2}(\1-N_{q_3})(\1-N_{q_4}),\\
&&h_{\alpha}^d=\sum_{\mathcal{N_\alpha}}N_{q_1}N_{q_2}N_{q_3}(\1-N_{q_4}),
\label{BCM}
\enano
where $\mathcal{N}_{\alpha}$ is the  neighborhood index of the cell $\alpha$\footnote{In the 1D case the neighborhood of a cell is made by the nearest-neighbor and next-nearest-neighbor cells}, and the sums in $h_{\alpha}^l, h_{\alpha}^d$ run on
every possible permutation of the indexes $q_1,q_2,q_3,q_4$ in $\mathcal{N}_{\alpha}$.
This Hamiltonian induces a dynamics similar to that induced by the standard rules in the classical game of life; in fact, the operators $h_{\alpha}^l$ and $h_{\alpha}^d$ count the densities in the neighboring cells of $\alpha$, and
 $h_{\alpha}^l$ ($h_{\alpha}^d$) is null if the sum of alive cells in the neighborhood of $\alpha$ is different from two (three). Densities in the cells are then computed during the time evolution through the expectation values of the number operators on the initial state. Our approach differs from the one proposed in \cite{BCM2012} not only for the different expression of the Hamiltonian which, in our case, contains a diffusion term of the population, but mainly because of our  application of the $(H,\rho)$-induced dynamics.

\section{Results}\label{sec:results}
In this Section, by using different tools, we perform  an in-depth analysis of the results that can be deduced out of our model. At first, we study the effects of the two parameters  entering the $(H,\rho)$-induced dynamics of the
\emph{QGoL}. These are  $\tau$, which defines the time range during which only the Hamiltonian--driven  evolution is
active before the application of the rule in \eqref{nr1}-\eqref{nr4}, and $\sigma$, which measures the deviation of the new rule with respect to the one originally given in (\ref{r1})-(\ref{r4}). Then, we
analyze the output of our model by means of both the spectral and  blob analysis.

All our simulations have been performed on a  two--dimensional square lattice of dimension $L^2$, with $L=33$, by
choosing  several initial configurations in which the state of each cell is initialized in a random way, with equal
probabilities to have value 0 or 1. Our results are compared with those deduced from the \emph{GoL} in order to
highlight the main effects due to the $(H,\rho)-$induced dynamics.

\subsection{The parameters $\tau$ and $\sigma$}
\label{sec:PTS}

The parameter $\tau$ defines the time range of the $(H,\rho)$--induced dynamics of the system  before the rules are applied. Obviously, for $\tau=0$  there is no
Hamiltonian--driven dynamics at all, and, therefore, if $\sigma=0$, we recover the  classical behavior of the \emph{GoL}. To study how the parameters $\tau$ and $\sigma$ modify the classical evolution, we first evaluate at the second generation ($K=2$) the following sort of mean $l_1$-error norm between the states  of the cells obtained by the quantum and the classical games of life:
\begin{equation}
\Delta_{QGoL}^{GoL}(\tau,\sigma)=\frac{1}{L^2}\sum_{\alpha=1}^{L^2}|n_{\alpha,2}-\mathring{n}_{\alpha,2}|,
\end{equation}
where $n_{\alpha,2}$ and $\mathring{n}_{\alpha,2}$ are the states in the cell $\alpha$ at the second generation for the \emph{QGoL} and the \emph{GoL}, respectively. Hence, $\Delta_{QGoL}^{GoL}(\tau,\sigma)=0$ when $n_{\alpha,2}=\mathring{n}_{\alpha,2}$ for all $\alpha$, \emph{i.e.}, when the \emph{QGoL}  and the \emph{GoL} actually coincide (at the second generation, and so at all generations).
To make our results more robust,  we have computed the distribution $\Delta_{QGoL}^{GoL}(\tau,\sigma)$  by averaging the differences obtained from 100 different random initial conditions for fixed $\tau$ and $\sigma$.
\begin{figure}
\begin{center}
\subfigure[Distribution of $\Delta_{QGoL}^{GoL}(\tau,\sigma)$]{\includegraphics[width=0.48\textwidth]{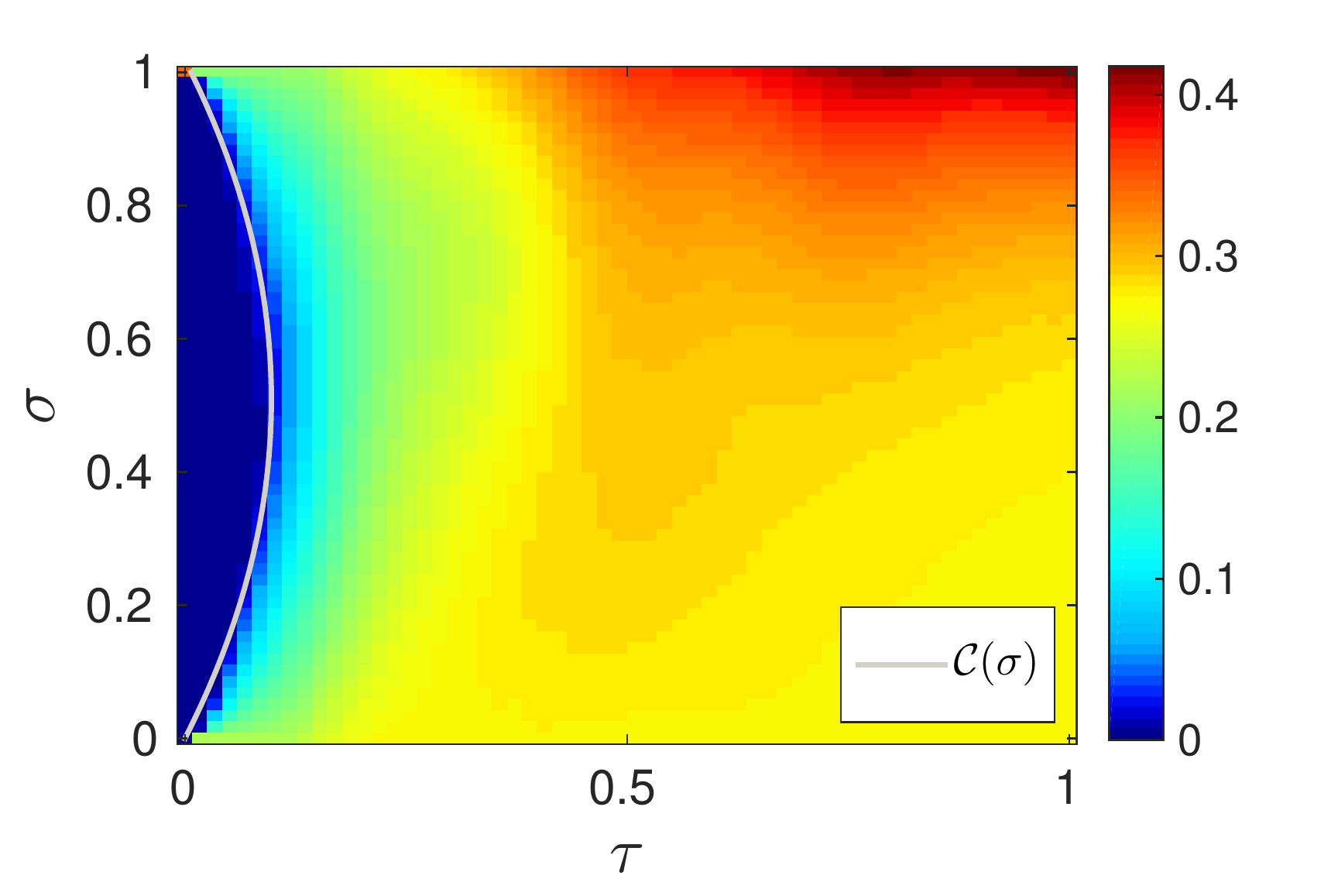}}
\subfigure[$\sigma_{min}(\tau)$ for $0\leq\tau\leq1$]{\includegraphics[width=0.48\textwidth]{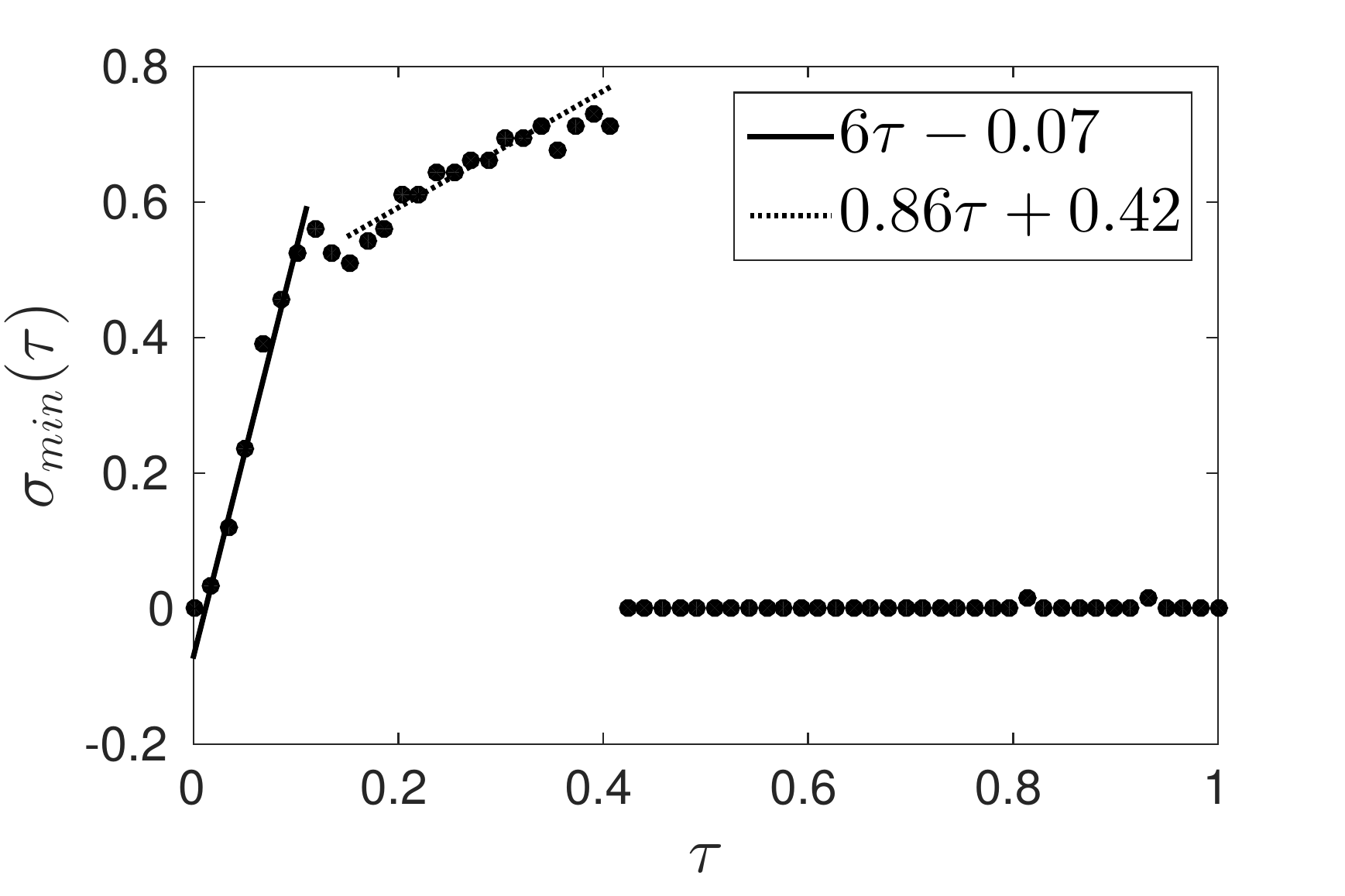}}	
\end{center}
\caption{\label{tau_mindelta}\footnotesize \textbf{(a)} The distribution $\Delta_{QGoL}^{GoL}(\tau,\sigma)$ of the $L_1$ error norms between the states of the cells obtained by the \emph{QGoL}  and \emph{GoL}. For a fixed $\sigma$, $\Delta_{QGoL}^{GoL}$ essentially increases with  the time $\tau$ in which the Hamiltonian--driven
evolution takes place. The dependence of  $\Delta_{QGoL}^{GoL}(\tau,\sigma)$ on $\tau$ is the following one:  for $\tau<0.4$,  $\Delta_{QGoL}^{GoL}(\tau,\sigma)$ increases as $\sigma$ approaches 0 or 1, while for $\tau>0.4$ the error increases with  $\sigma$. The white curve is
the quadratic curve $\mathcal{C}(\sigma)=-0.337\sigma^2+0.384\sigma$ approximating the contour level $\Delta_{QGoL}^{GoL}(\tau=0.1,\sigma=0.5)=0.02$.
  \textbf{(b)} The minimum $\sigma_{min}(\tau)$ of $\Delta_{QGoL}^{GoL}(\tau,\cdot)$ for a fixed $\tau$. Different linear growth rates are visible for three $\tau$ ranges: for $\tau>0.4$ $\sigma_{min}(\tau)\approx0$ which is explicative of the fact that there is a phase transition  for $\tau\simeq0.4$.     }
\end{figure}
The distribution $\Delta_{QGoL}^{GoL}(\tau,\sigma)$ is shown in Fig.~\ref{tau_mindelta}(a) for $0\leq\tau,\sigma\leq1$. The way in which $\tau$ affects  $\Delta_{QGoL}^{GoL}$
is clear: $\Delta_{QGoL}^{GoL}$ increases with $\tau$. This is in agreement with the fact that $\tau>0$ corresponds to an effect which is absent in the classical situation, since, in this case, no time evolution exists at all. In the {\em GoL}, in fact, the rule only, applied again and again, creates the different generations. The dependence of $\Delta_{QGoL}^{GoL}(\tau,\sigma)$ on $\sigma$ is much richer, since it is also related to the value of $\tau$. For $\tau<0.4$,  $\Delta_{QGoL}^{GoL}(\tau,\sigma)$  increases as $\sigma$ approaches 0 or 1, while it decreases for intermediate values of $\sigma$.
On the other hand, for $\tau>0.4$ the error increases with $\tau$ and with $\sigma$, taking its minimum value for $\sigma=0$.
This is essentially what one could expect, since larger values of $\tau$ and $\sigma$ represent bigger differences from the classical situation, which is exactly recovered if $\sigma=\tau=0$. Nevertheless, it is interesting to notice that, for $\tau<0.4$, $\sigma=0$, we have a maximum value for $\Delta_{QGoL}^{GoL}(\tau,\sigma)$. This suggests that, even if $\sigma=0$ (so that the new rules and the classical ones do coincide), the time evolution of the system driven by $H$ is already enough to significantly modify the behavior of the system.

Our numerical simulations also suggest that, in general, for a fixed $\tau$, there is  always a value $\sigma_{min}(\tau)$ for which  $\Delta_{QGoL}^{GoL}(\tau,\sigma_{min}(\tau))$ reaches a minimum. In particular, for very small values of $\tau$,  there exist ranges of parameter $\sigma$ for which $\Delta_{QGoL}^{GoL}(\tau,\sigma)$ is vanishing, so that the \emph{QGoL} and \emph{GoL} dynamics coincide: for instance, for $\tau=0.01$, we obtain  $\Delta_{QGoL}^{GoL}(\tau,\sigma)=0$ for $0.009<\sigma<0.99$. This fact suggests, once again, that the role of the action of $H$ is more relevant than the change in the rule $\rho$ (\emph{i.e.}, the passage from the classical rule to the new one). For later convenience, if for a fixed $\tau$ we have a range of minima $[\sigma_1,\sigma_2]$ of $\Delta_{QGoL}^{GoL}(\tau,\sigma)$, then we fix $\sigma_{min}(\tau)=\sigma_1$. In Fig.~\ref{tau_mindelta}(b) we plot $\sigma_{min}(\tau)$, and a piecewise linear behavior with three different  slopes is visible; $\sigma_{min}(\tau)$ appears increasing for $\tau\leq0.4$. In particular, for $\tau\leq0.1$,  $\sigma_{min}(\tau)$ has a linear growth rate of $6$, while for $0.1<\tau\leq0.4$ the linear growth rate is much lower, close to $0.87$. For $\tau>0.4$ $\sigma_{min}(\tau)\approx0$. It looks like a phase transition for $\tau\simeq0.4$, but, so far, the reason for such a transition is not clear. This strange behavior suggests a deeper analysis of $\sigma_{min}(\tau)$, which is postponed to a future paper.

In Fig.~\ref{tau_mindelta}(a), we also show the quadratic curve
$\mathcal{C}(\sigma)=-0.337\sigma^2+0.384\sigma$ approximating the contour level
$\Delta_{QGoL}^{GoL}(\tau=0.1,\sigma=0.5)=0.02$. This contour level surrounds the region
$\tau<\mathcal{C}(\sigma)$ in which $\Delta_{QGoL}^{GoL}$ has its lowest values, and, as we shall see in
the Appendix, it allows to characterize the region  of $\tau$ and $\sigma$ where the periodicity of a periodic
orbit of the \emph{QGoL} case differs from the \emph{GoL} case.

\subsection{Spectral analysis of the \emph{QGoL}}\label{Spectral}

Here we perform a statistical study of the \emph{QGoL} by using the classical tools of the spectral analysis. In particular, for a fixed cell $\alpha$,  the Fourier transform of its state $n_{\alpha,k}$  at the various generations $k = 0,\ldots , T- 1$ is
given by
\be
\tilde n_{\alpha,k} (f)=\frac{1}{T}\sum_{t=0}^{T-1} n_{\alpha,k} \textnormal{exp}\left(-i\frac{2\pi tf}{T}\right),
\en
and the Fourier power spectrum  is defined as
\be
S(f)=\sum_{\alpha=1}^{L^2}| \tilde n_{\alpha,k}(f) |^2.
\label{spectrum}
\en
Moreover, we also consider  the density of alive cells at generation $k$--th, defined as
\be
D^k=\frac{1}{L^2}\sum_{\alpha=1}^{L^2} n_{\alpha,k}.
\label{density}
\en
Roughly speaking, the power spectrum  $S(f)$  gives information on the frequencies excited due to the possible presence of an equilibrium cycle solution of period  $T/f$. The density of alive cells $D^k$ is the ratio of alive cells for each generation $k$, and stationary or periodic behavior of $D^k$ gives information about possible periodicity of the solution (in the sense of Definition \ref{def2}).

\begin{figure}
\begin{center}
\subfigure[Power spectrum for the \emph{GoL}, and the \emph{QGoL} for $\tau=0.1$]{\includegraphics[width=0.48\textwidth]{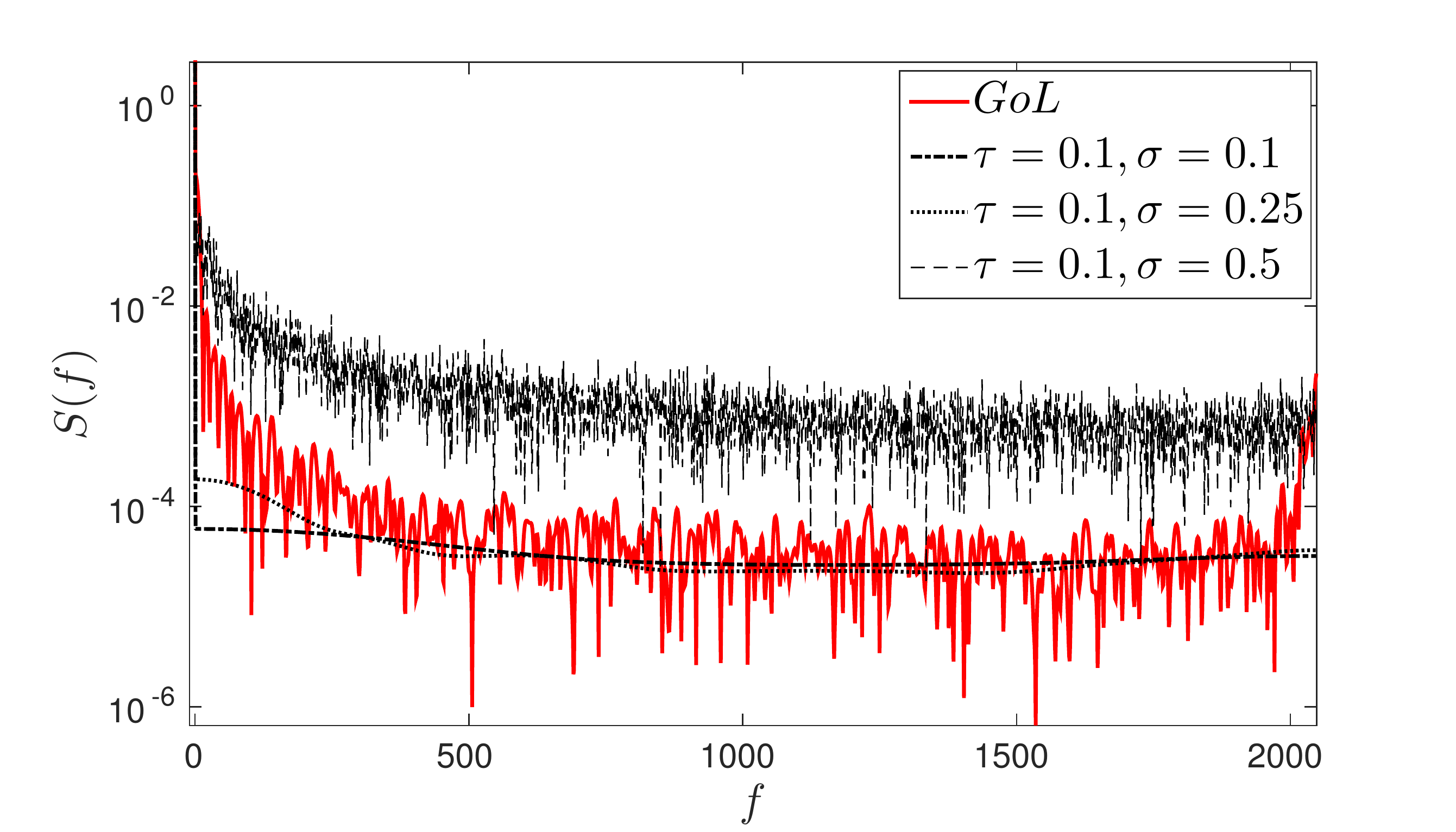} 	 \label{spectrum01}}
\subfigure[Density of alive cell for the \emph{GoL}, and the \emph{QGoL} for $\tau=0.1$]{\includegraphics[width=0.48\textwidth]{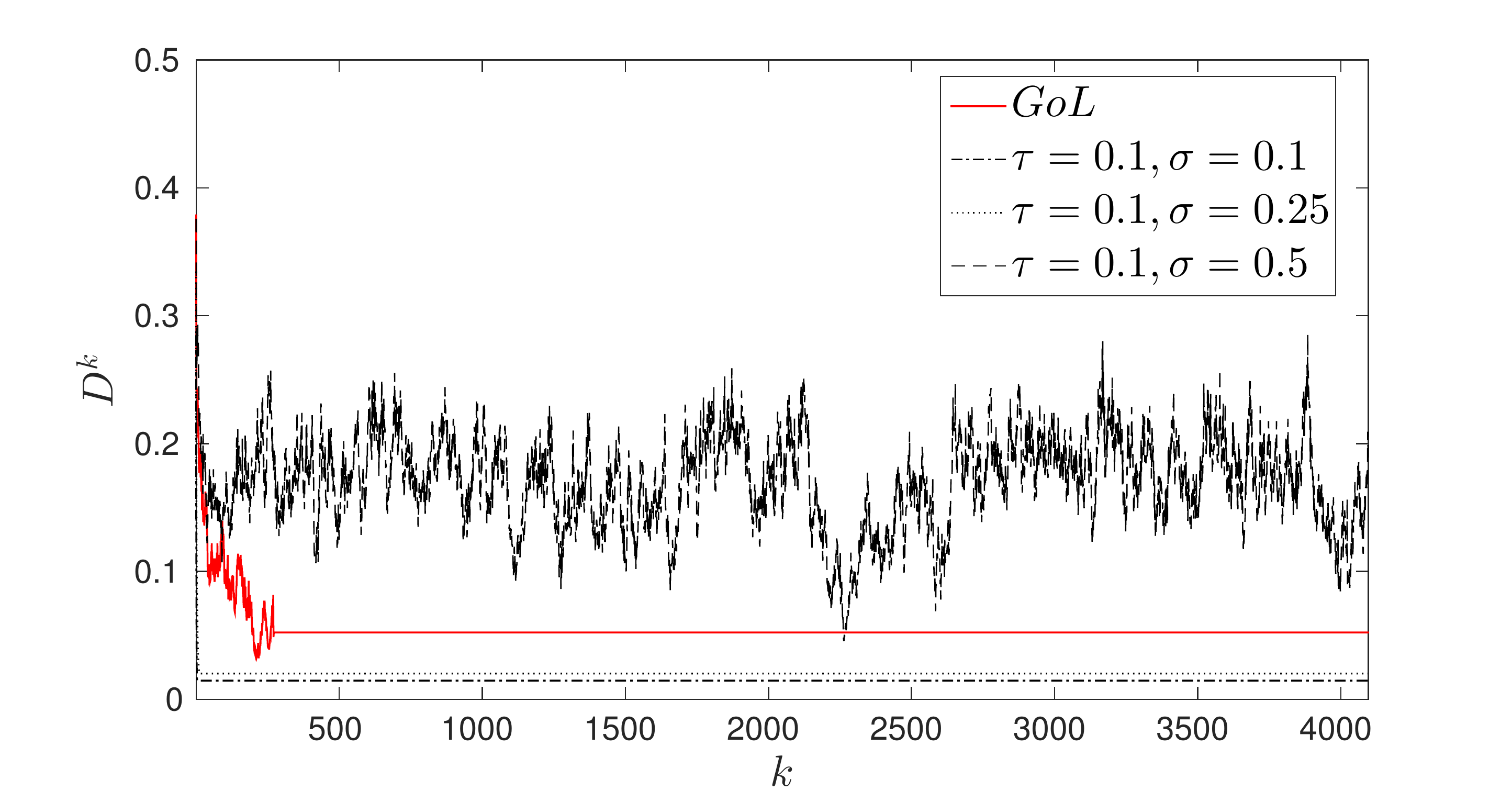}	 \label{density01}}
\end{center}
\caption{\footnotesize \textbf{(a)} The power spectrum for the \emph{GoL} and the \emph{QGoL} for $\tau=0.1$ and various $\sigma$ with a random initial condition and $T=4096$ generations. The \emph{GoL} exhibits a $1/f$ power spectrum, and has a 2--equilibrium cycle solution after a transient of 277 generations (see the density of the alive cell in \textbf{(b)}), leading to the peak in the frequency $f=2048$ of the spectrum. The \emph{QGoL} spectrum exhibits for $\sigma=0.1,0.25$ a low power density as a consequence of a 1--equilibrium cycle solution obtained after very few generation, while for $\sigma=0.5$ the spectrum has a $1/f^{0.15}$ behavior with all frequencies excited due to non periodic solution (see \textbf{b)}). }
\end{figure}

It is well known that  the \emph{GoL} has $1/f$ noise \cite{NYH}, \emph{i.e.}, its power spectrum  behaves like
$1/f$ at low frequencies, and in general cellular automata can have a power spectrum of the kind $f^{\alpha}$
\cite{NYH2}; $1/f$ noise can be observed in a wide
variety of phenomena such as the voltage of vacuum tubes, the rate of
traffic flow, and the loudness of music. According to \cite{kes}, a system showing a $1/f$ power spectrum
is such that its current state is influenced by the history of the system itself. The presence of the $1/f$
behavior of the power spectrum has also been found  in \cite{LAPM}, in the case of an asynchronous version of the \emph{GoL}. Evidence of the $1/f$ noise is given in
Fig.~\ref{spectrum01}, where the power spectrum for the \emph{GoL} is shown for an initial random condition and $T=4096$ generations\footnote{Similar results arise also for other initial random conditions.}.
For this initial condition, according to Definition \ref{def2}.3 and considering $\tau=0$, $\sigma=0$,  we have obtained a 2-equilibrium cycle after a transient of 277 generations (see the density of alive cells in
Fig.~\ref{density01}). Hence, in the power spectrum, there is a final peak  at the frequency $2048$, due to the fact that a 2-equilibrium cycle solution is a periodic orbit with period $\Omega=2$ giving strength to the frequency $T/\Omega=2048$.  By fitting the spectrum $S(f)$ with a function $Cf^{\alpha}$ with a least square method, we obtain in the range $f=1,\ldots,2000$ the values $C=0.4$, $\alpha=-1.033$, consistent with the predicted $1/f$ of the power spectrum. If we consider  the \emph{QGoL} case, in the same
Fig.~\ref{spectrum01} there are shown the power spectra for $\tau=0.1$ and various
values of $\sigma$. As for as the values
$\sigma=0.1,0.25$ are concerned, the spectrum  is characterized by low power density at almost all frequencies with a peak at the first frequency $f=0$: this is due to the circumstance that, after an initial transient, the whole system stabilizes to a 1-equilibrium cycle solution  after very few generations. In fact, for the same initial random condition used for \emph{GoL}, the \emph{QGoL} stabilizes to a
$4$--equilibrium  and $6$--equilibrium cycle solution for $\sigma=0.1$ and $\sigma=0.25$, respectively. For higher values of $\sigma$ ($\sigma=0.5$), the spectrum has almost all frequencies excited with a clear low power behavior for small frequencies. The fitting with the function
 $Cf^{\alpha}$ for $f=0,\ldots,200$ returns $C=0.45$, $\alpha=-0.15$ for $\sigma=0.5$. The circumstance that almost all the frequencies are excited, with decreasing amplitude, means that the  solution does not show (at least for the number of generations considered) any periodicity or equilibrium. However, it is important to stress that, because of the finite dimensionality of our system and of the finite number of the possible states of each cell, each initial condition necessarily generates a periodic solution (the worst possible case is that an initial state $\phi_{\mathbf{n}^0}$ returns in itself after $2^{L^2}-1$ generations). This suggests that, in order to detect the periodic structure in the solution, we need to consider a larger number of generations.

\begin{figure}
\begin{center}
\subfigure[Power spectrum for $\tau=0.25$]{\includegraphics[width=0.48\textwidth]{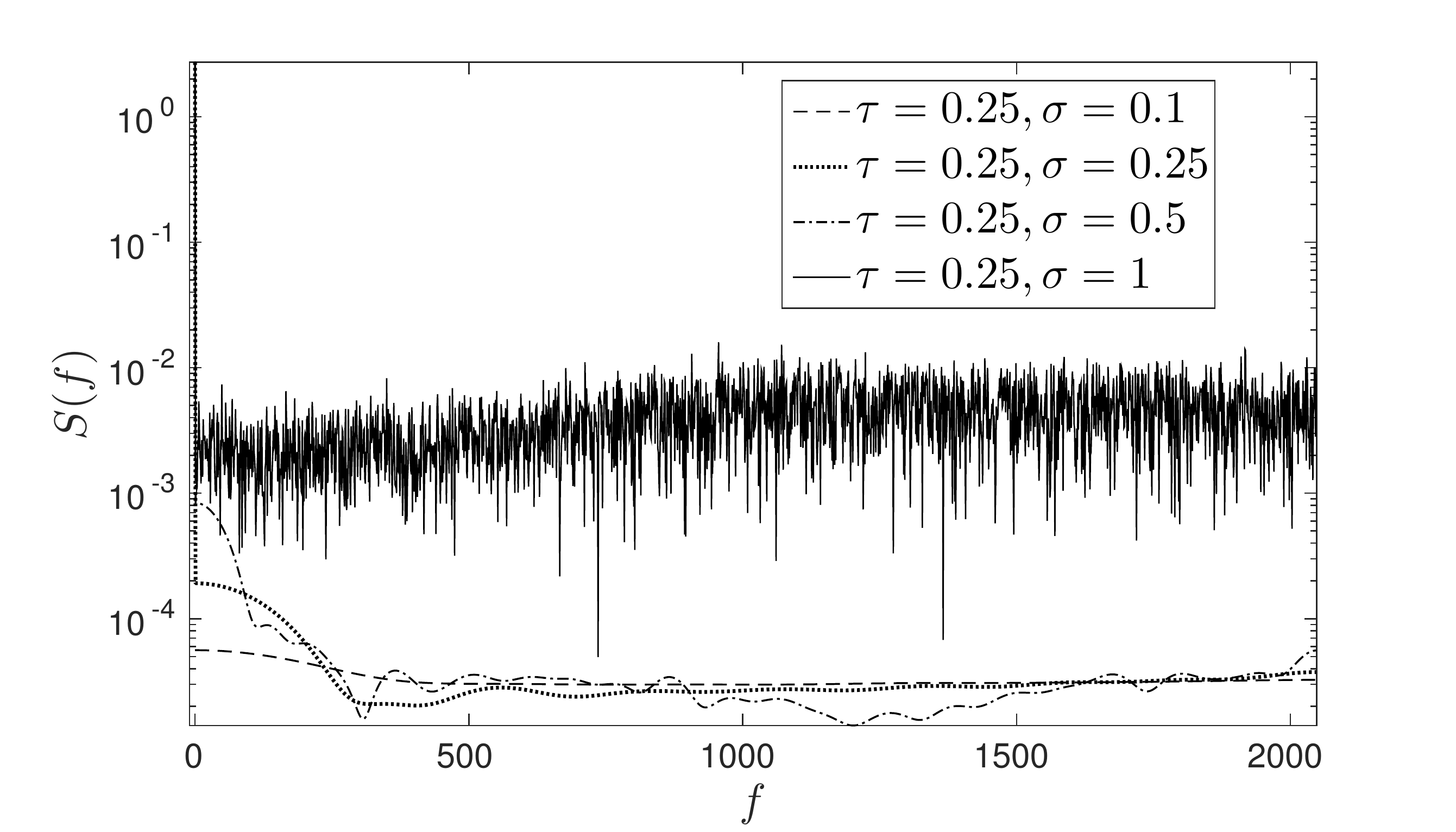} 	 \label{spectrum025}}
\subfigure[Density of alive cell for $\tau=0.25$]{\includegraphics[width=0.48\textwidth]{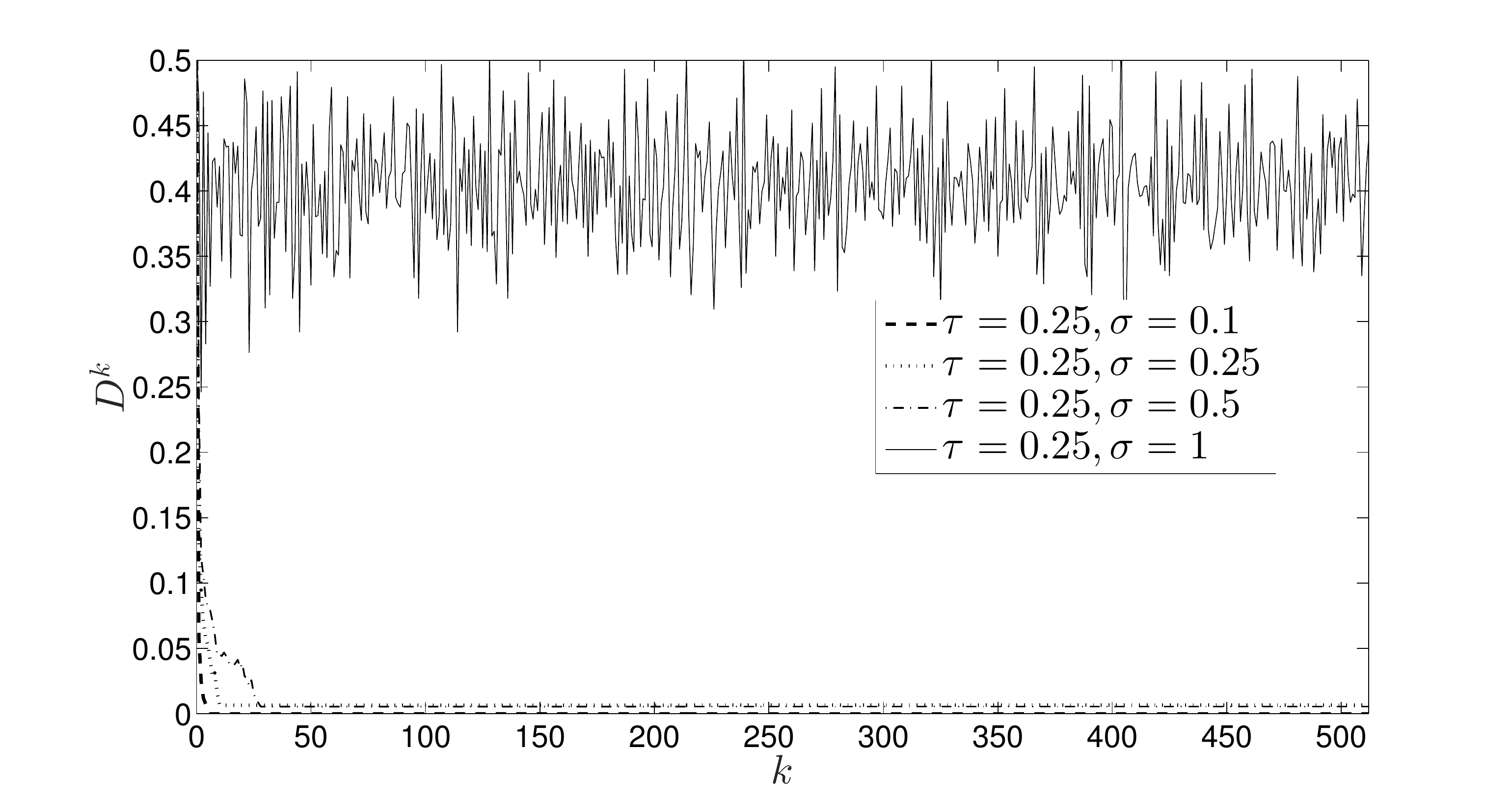}	 \label{density025}}
\end{center}
\caption{\footnotesize\textbf{a)} Same as Fig. \ref{spectrum01} for the \emph{QGoL} case and $\tau=0.25$. Also in this case for $\sigma\leq0.5$ the power spectrum has low power density with a peak at the frequency $f=0$ due to a 1--equilibrium solution while for $\sigma>0.5$ the spectrum has a noisy behavior  with a high number of alive density cell (see figure \textbf{b)}). }	 
\end{figure}

\begin{figure}
\begin{center}
\subfigure[Power spectrum for $\tau=0.5$]{\includegraphics[width=0.48\textwidth]{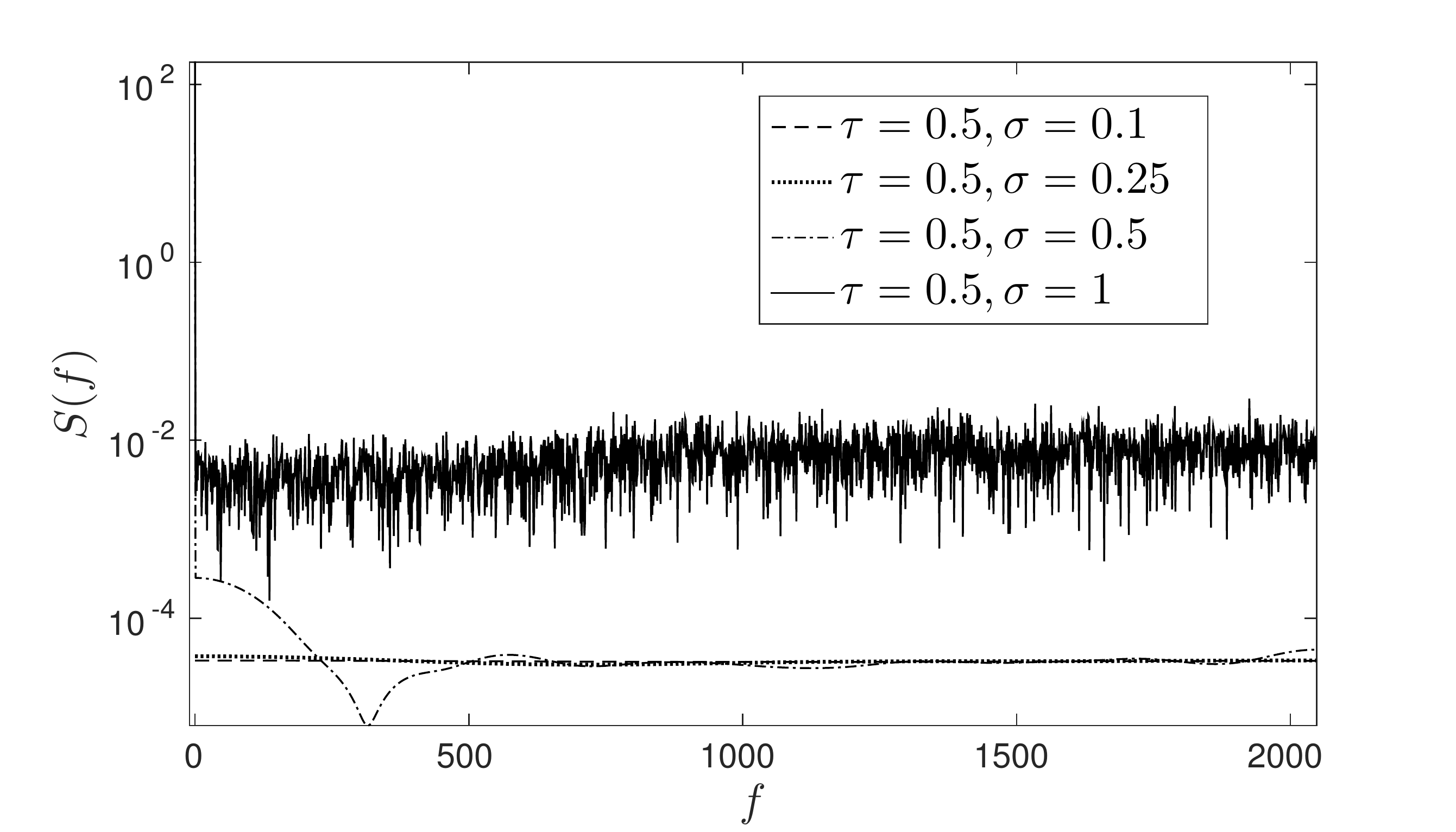} 	 \label{spectrum05}}
\subfigure[Density of alive cell for $\tau=0.5$]{\includegraphics[width=0.48\textwidth]{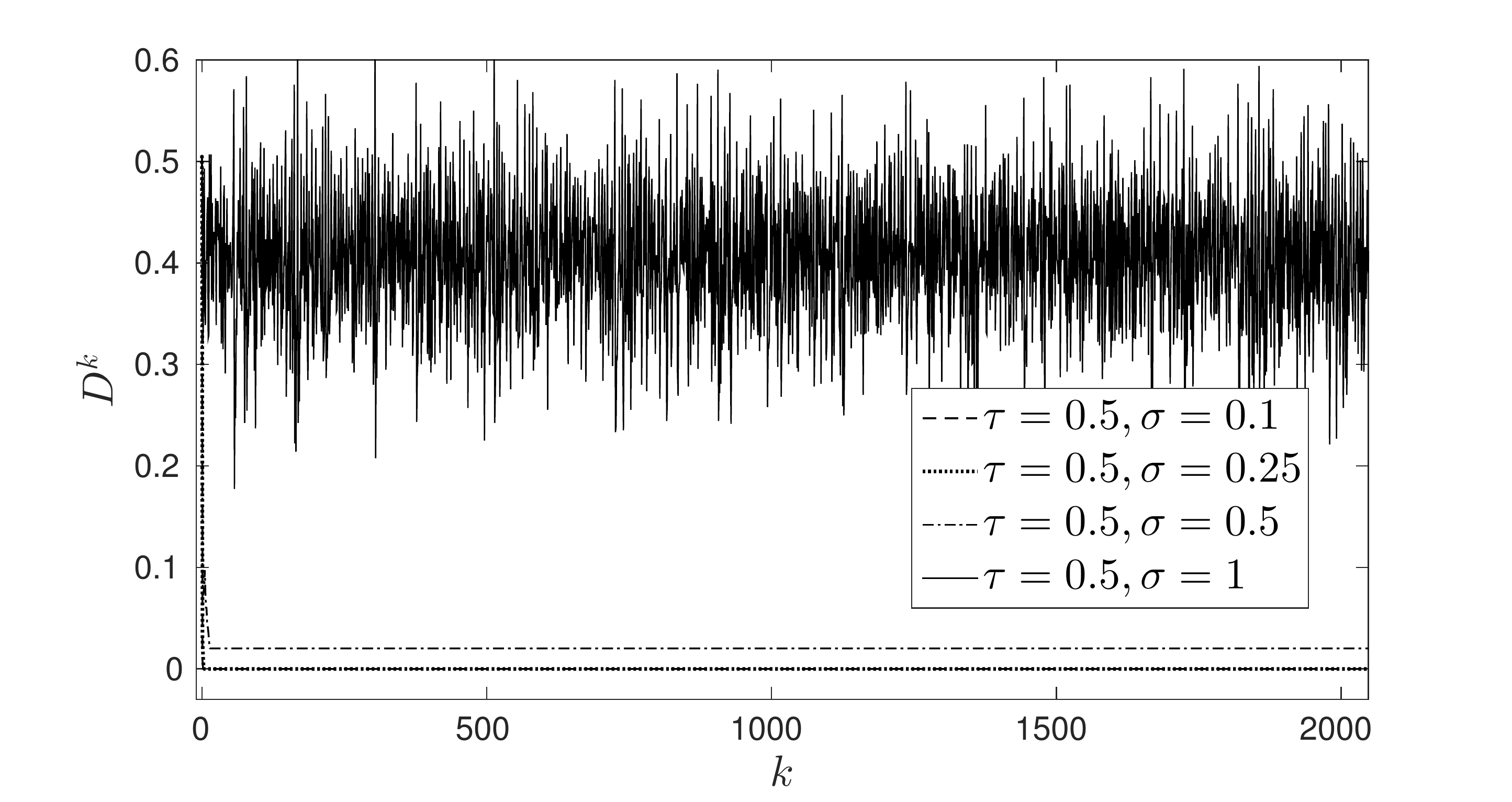}	 \label{density05}}
\end{center}
\caption{\footnotesize\textbf{a)} Same as Fig. \ref{spectrum01} for the \emph{QGoL} case and $\tau=0.5$. Results are similar to those obtained for the case $\tau=0.25$ (see Fig.\ref{spectrum025})}	
\end{figure}

For increasing values of $\tau$ ($\tau=0.25,0.5$),  the situation does not change  for $\sigma\leq0.5$, since all the power spectra are similar to those observed for the case $\tau=0.1$, $\sigma=0.1,0.25$,  with a low power density at almost all frequencies, and a peak at the first frequency $f=0$; thus, there is a 1--equilibrium cycle solution after few iterations, as shown by the density of alive cells in Figs.~\ref{density025}-\ref{density05}.
For $\tau=0.25,0.5$ and $\sigma=1$ we have a peak at $f=0$, and the remaining frequencies excited with almost the same amplitude similar to a ``noise'' signal, meaning that the evolution is virtually orderless with an high number of alive cells in each generation (see Figs.~\ref{density025}-\ref{density05}): still in this case, for the number of generations we have considered, we have not obtained an equilibrium periodic solution.


Comparing these results with those of the \emph{GoL}, we may observe that for small values of $\tau$ and $\sigma$,  corresponding to small expected variations from the classical situation, we recover equilibrium solutions with periods  which are smaller than those found for $\tau=\sigma=0$.  On the other hand, for $\tau$ and $\sigma$ close to 1, and the number of generations considered in our numerical simulations, the equilibrium is not observed, so that the situation is really different from the one corresponding to the classical \emph{GoL}. However, as already stated,  such an equilibrium solution must exist also in our setting, even if the transient period, for large values of $\tau$ and $\sigma$, may be so long that the equilibrium is not observed during the numerical tests.
To give an insight on the equilibrium cycle solution formed in our simulations, we will consider in the Appendix a case study ($L=5$), and characterize all possible equilibria and their transients.

\subsection{Blob analysis}\label{blobanalysis}

This Section deals with the so-called blob analysis \cite{blob} of the generations of both \emph{GoL} and \emph{QGoL}. More precisely, each generation, that, as stated, is essentially a distribution of 0 (for dead cells) and 1 (for alive cells) over a lattice, can be represented as a binary image. In particular, we performed the analysis of the
8--connected largest alive components in the binary images (our {\em blobs}) corresponding to the states of the system after each generation of various simulations of the \emph{GoL} and the \emph{QGoL}. By means of a forward scan of the lattice, each time an alive cell is encountered, we  use it as a seed for the reconstruction of the binary large object of alive neighboring cells it belongs to.

For different choices of the parameters $\tau$ and $\sigma$, the analysis of the blobs detected during the evolution of the system has been carried out up to a stationary or periodic behavior of the patterns, with particular focus on the following properties:
\begin{itemize}
\item total number of blobs for configuration;
\item area of each blob;
\item perimeter of each blob;
\item centroid of each blob;
\item centroid of the whole configuration.
\end{itemize}
Area, perimeter and circularity are features used in shape analysis. The area of an alive connected region can be accurately estimated by counting the number of the cells of value $1$ of the region. To obtain a good perimeter estimator, a contour following procedure using distances in taxicab geometry has been performed.
To compute the circularities, the ratio between perimeters and areas of the various regions has been simply considered.

\begin{figure}[h]
\begin{center}
\subfigure[\label{living01} \emph{GoL} producing a 2--equilibrium cycle solution from step 594.]
{\includegraphics[width=0.48\textwidth]{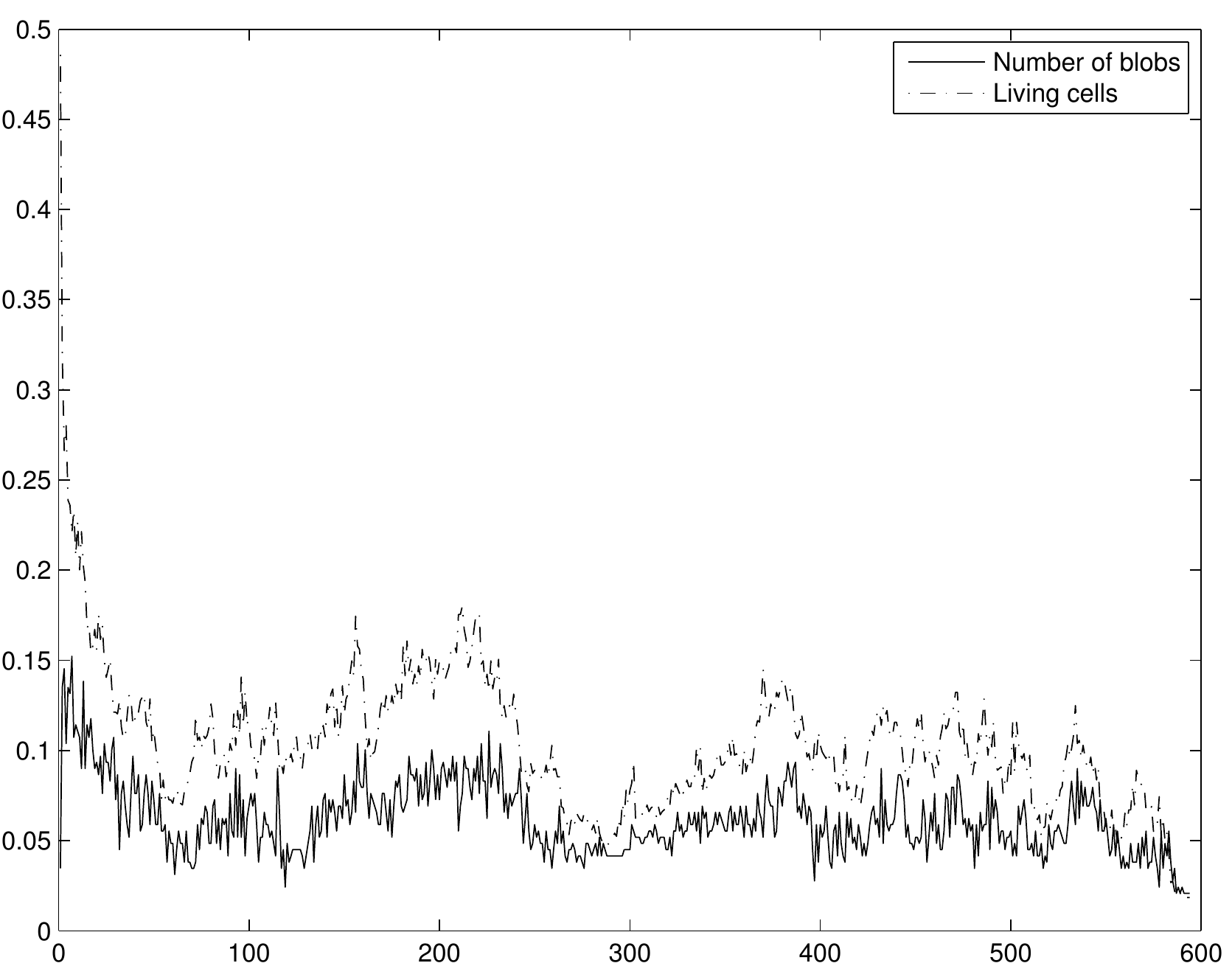}}
\subfigure[\label{living02} \emph{QGoL}, for $\tau=0.25$ and $\sigma=0.1$, stable from step 9.]
{\includegraphics[width=0.48\textwidth]{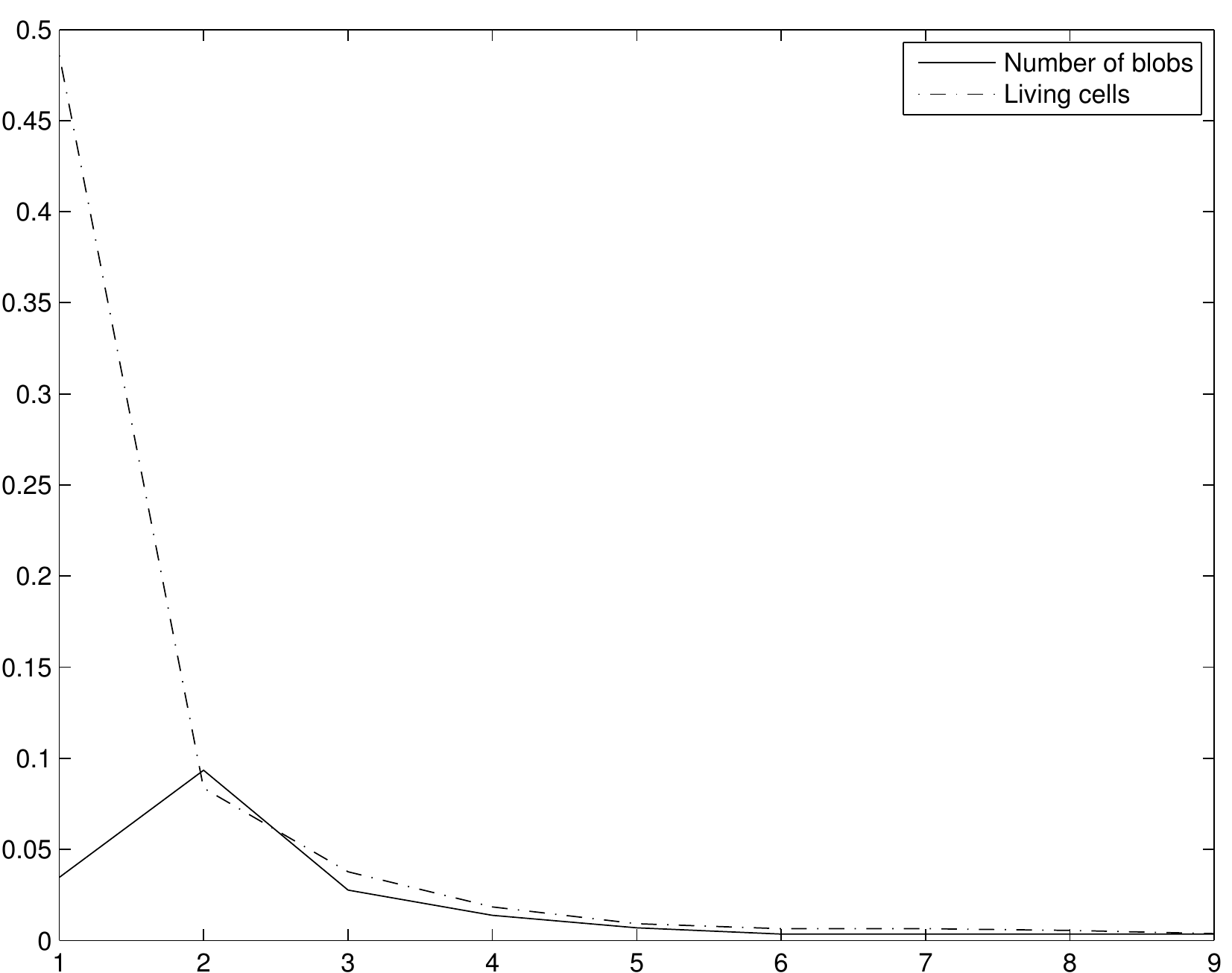}}
\end{center}
\caption{\footnotesize Number of living cells and blobs for the \emph{GoL}, \ref{living01}, and the \emph{QGoL}, \ref{living02}, for parameters $\tau=0.25$, $\sigma=0.1$, normalized with respect to the dimension of the lattice and the maximum number of connected objects in it, respectively. The evolution trends of the amount of alive cells and connected regions are similar either in the classical or in the quantum game of life. For the same initial condition, the \emph{QGoL} stabilizes to a
1--equilibrium cycle solution after very few generations, while the corresponding \emph{GoL} generates a cyclic solution of period 2 from step 594.}
\end{figure}

\begin{figure}[h]
\begin{center}
\subfigure[\label{living03} \emph{GoL} producing a 2--equilibrium cycle solution from step 245.]
{\includegraphics[width=0.48\textwidth]{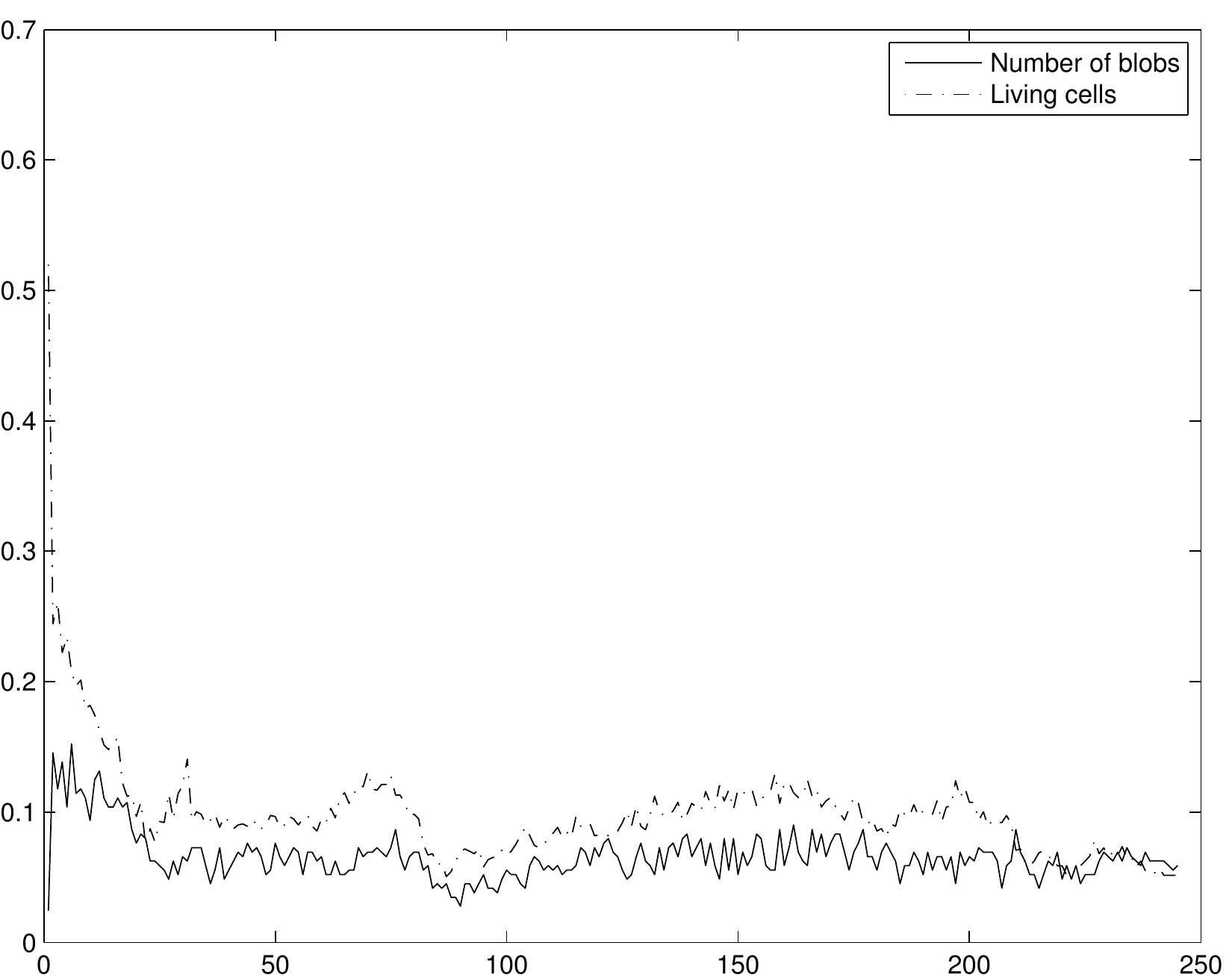}}
\quad
\subfigure[\label{living04} \emph{QGoL}, for $\tau=0.5$ and $\sigma=1$, not yet stable at step 4096.]
{\includegraphics[width=0.48\textwidth]{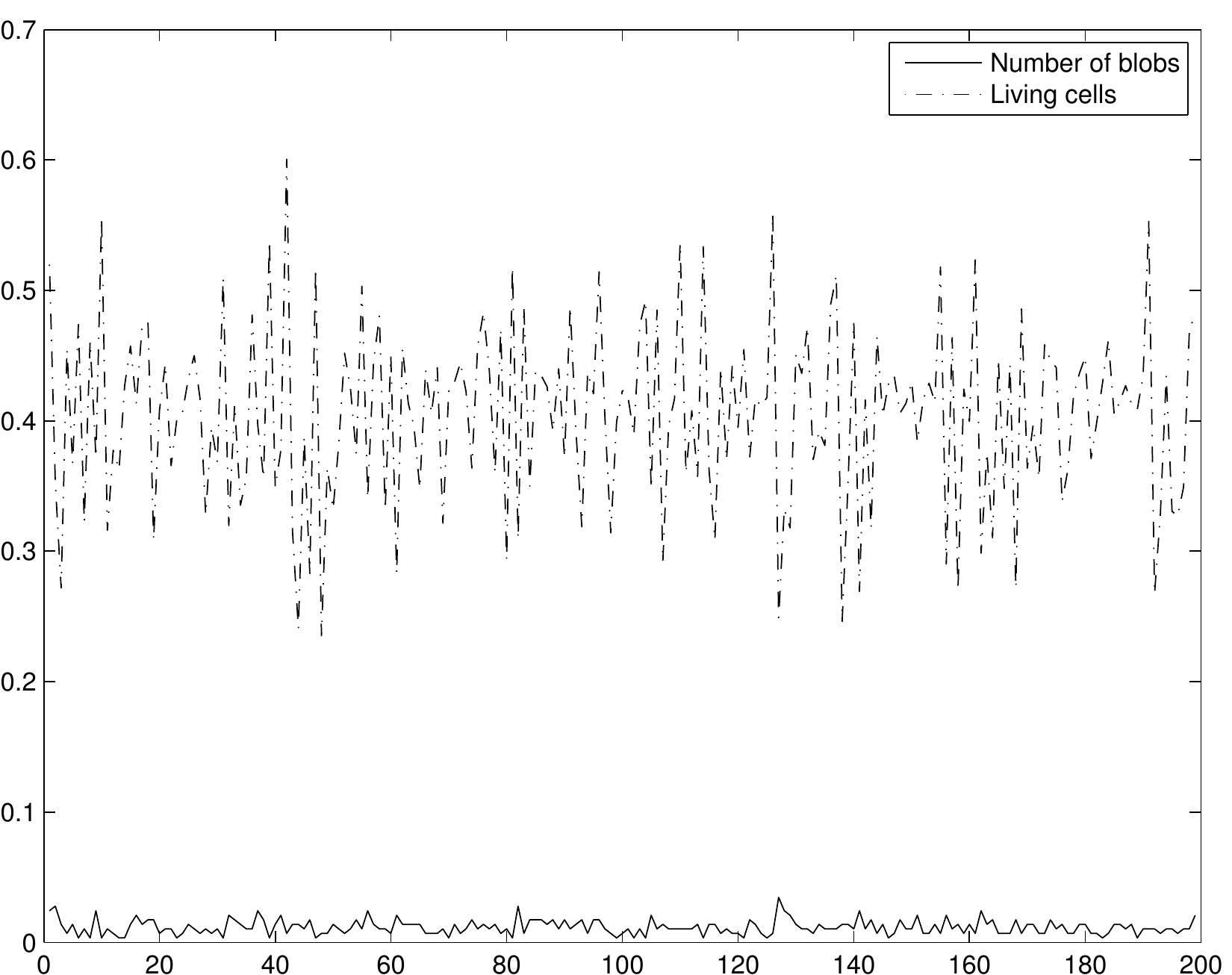}}
\end{center}
\caption{\footnotesize Number of living cells and blobs for the \emph{GoL}, \ref{living03}, and the \emph{QGoL}, \ref{living04}, for parameters $\tau=0.5$, $\sigma=1$, normalized with respect to the dimension of the lattice and the maximum number of connected objects in it, respectively. The evolution trends of the amount of alive cells and connected regions are similar either in the classical or in the quantum game of life. For the same initial condition, the \emph{GoL} generates a 2--equilibrium cycle solution after 245 generations, while the corresponding \emph{QGoL} requires a significantly high number of generations to get an equilibrium periodic solution.}
\end{figure}

Comparing the evolution of the number of alive cells and the total amount of connected regions, normalized with respect to the size of the lattice and the largest possible number of its connected components, respectively, the graphs plotted in Figs.~\ref{living01}--\ref{living02} and in Figs.~\ref{living03}--\ref{living04} reveal similar trends for the two curves, without strong fluctuations after few steps either in the quantum case or in the classical one. As already discussed in Subsection~\ref{Spectral}, on varying the parameter $\sigma$, the evolutions of \emph{QGoL} are characterized by the achievement of stability within the first few steps (for $\sigma$ less than $0.5$), unlike the corresponding classical evolutions; on the contrary, for values of $\sigma$ greater than or equal to $0.5$, a significant delay in reaching stable configurations compared to the classical case, which on average stabilize at most within about a thousand steps, is observed.

\begin{figure}
\begin{center}
\subfigure[\label{circularities01} Blobs' circularity parameters for the \emph{GoL}.]
{\includegraphics[width=0.48\textwidth]{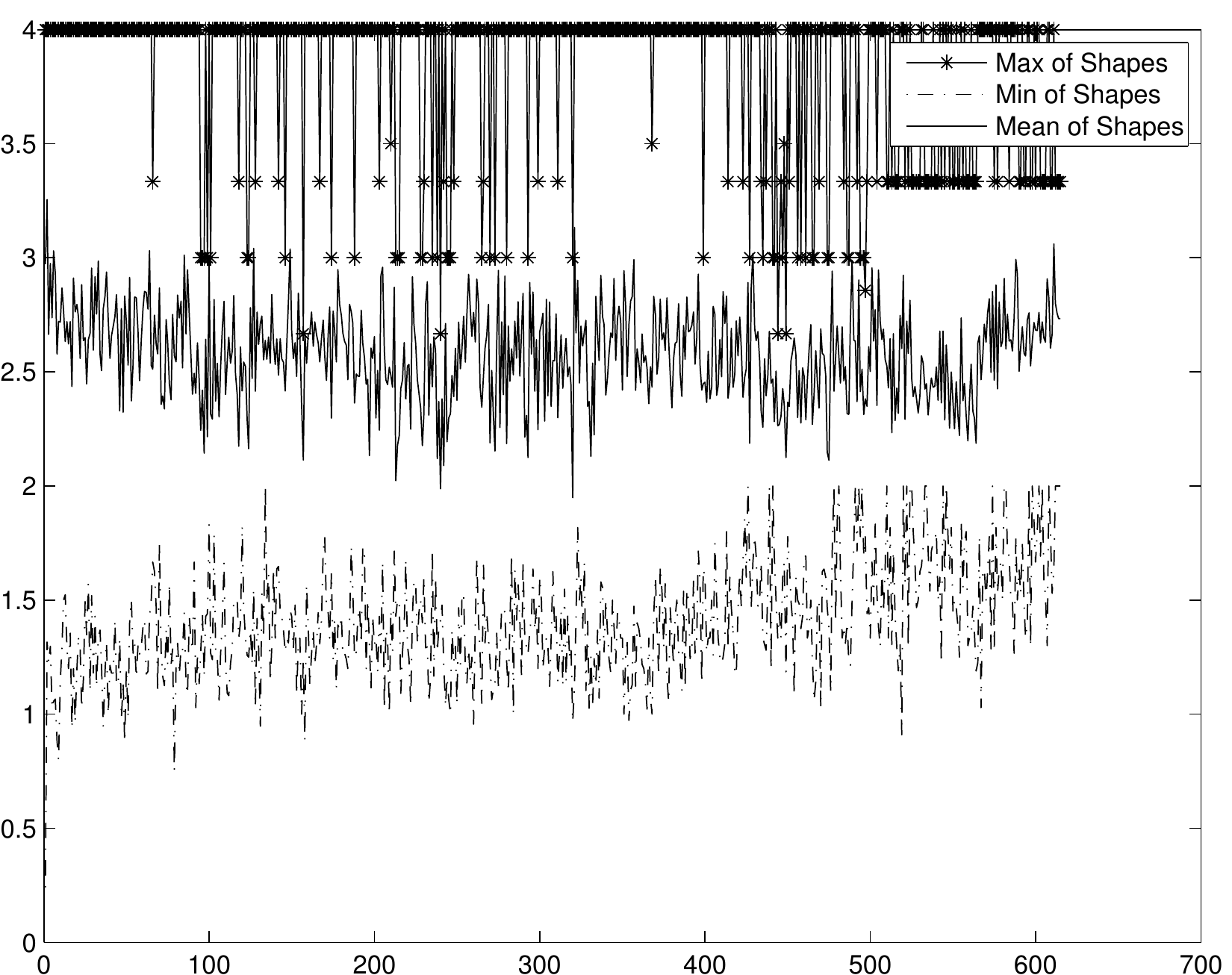}}
\subfigure[\label{circularities02} Blobs' circularity parameters for the \emph{QGoL}, for $\tau=0.1$ and $\sigma=0.25$.]
{\includegraphics[width=0.48\textwidth]{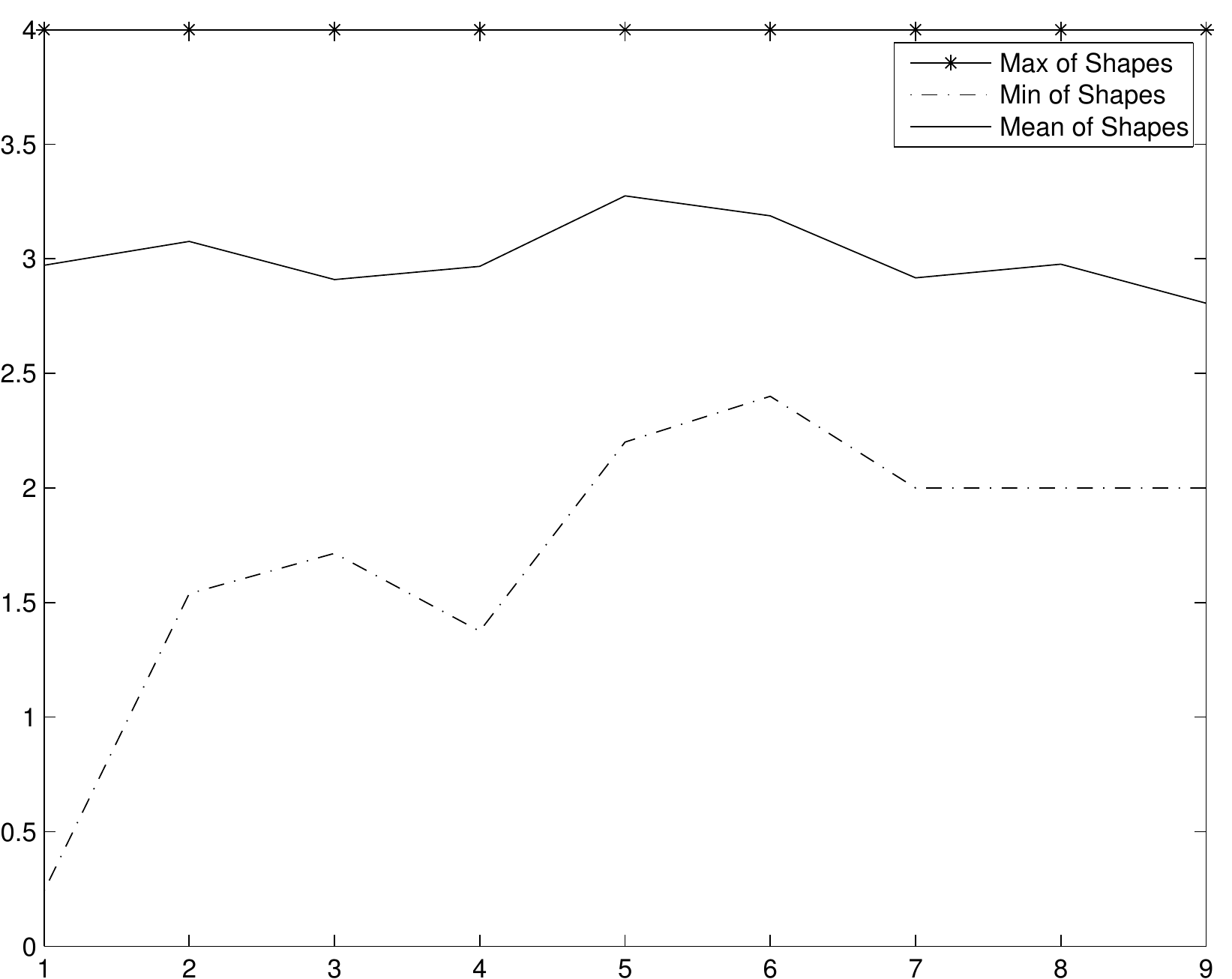}}
\end{center}
\caption{\footnotesize Trends of the maximum, minimum and average value of the circularity parameters of the polygons corresponding to the blobs at each generation of the \emph{GoL}, \ref{circularities01}, and the \emph{QGoL}, \ref{circularities02}, for parameters $\tau=0.1$ and $\sigma=0.25$. The fact that the maximum of the circularities of the blobs equals 4 at each step of the \emph{QGoL} attests the presence of single isolated alive cells or groups of living cells with at most one vertex in common during all the quantum evolution. Blobs with such a shape are not always detected in the classical case.}
\end{figure}

\begin{figure}
\begin{center}
\subfigure[\label{circularities03} Blobs circularity parameters for the \emph{GoL}.]
{\includegraphics[width=0.48\textwidth]{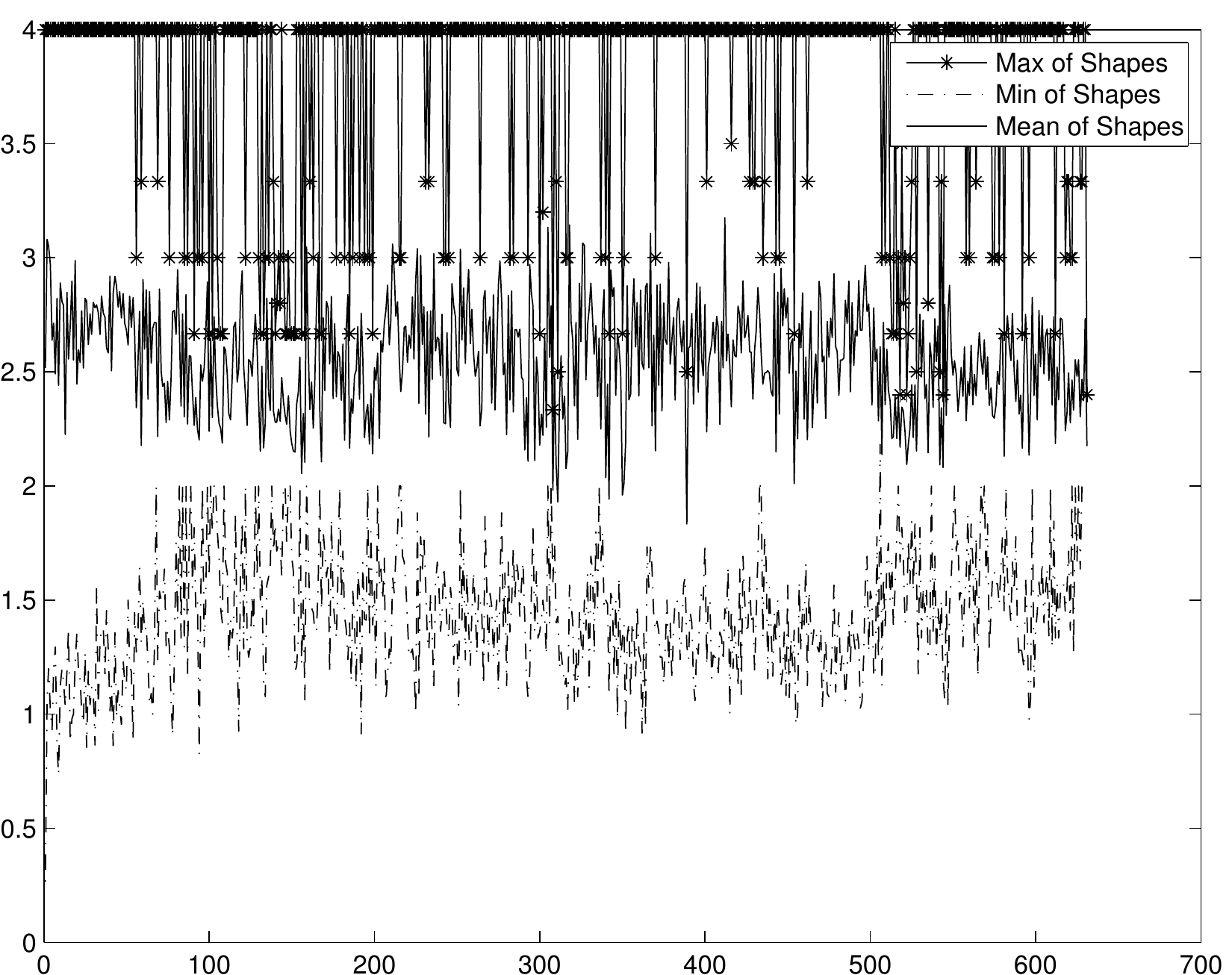}}
\subfigure[\label{circularities04} Blobs circularity parameters for the \emph{QGoL}, for $\tau=0.5$ and $\sigma=0.25$.]
{\includegraphics[width=0.48\textwidth]{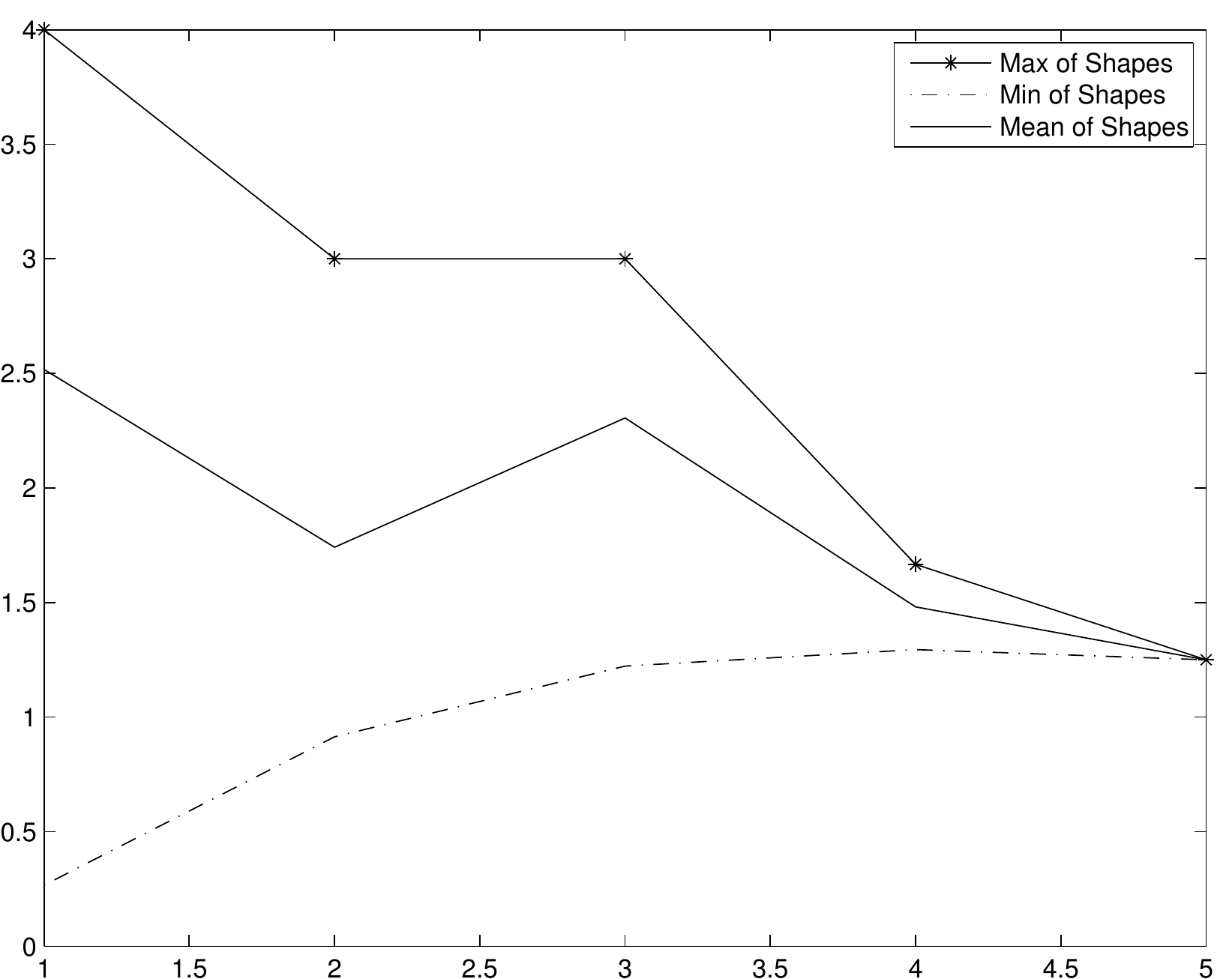}}
\end{center}
\caption{\footnotesize Trends of the maximum, minimum and average value of the circularity parameters of the polygons corresponding to the blobs at each generation of the \emph{GoL}, \ref{circularities03}, and the \emph{QGoL}, \ref{circularities04}, for parameters $\tau=0.5$ and $\sigma=0.25$. For values of $\tau$ greater than $0.1$, the circularity parameters of the blobs flatten to the average value.}
\end{figure}

The trends of the maximum, minimum and average value of the circularity parameters of the polygons corresponding to the blobs in different configurations provide a measure of how the shape of these connected regions deviates from the square shape, for which this value equals 4 divided by the number of neighboring cells. The maximum value of the circularity is reached in the case of single isolated alive cells or groups of living cells with at most one vertex in common. In the quantum case with $\tau = 0.1$, these types of connected components appear almost always, unlike the corresponding configurations in the classical case (see Figs.~\ref{circularities01}--\ref{circularities02}). In any case, as expected for a very short quantum interaction, the general trend of the curves for the shape parameters looks similar both for the \emph{GoL} and the \emph{QGoL}. As $\tau$ increases, however, as shown in Figs.~\ref{circularities03}--\ref{circularities04}, the values corresponding to the shapes of the connected components tend to the average values.

\begin{figure}[h]
\begin{center}
\subfigure[\label{centroids01} Centroid occurrences for the \emph{GoL}.]
{\includegraphics[width=0.48\textwidth]{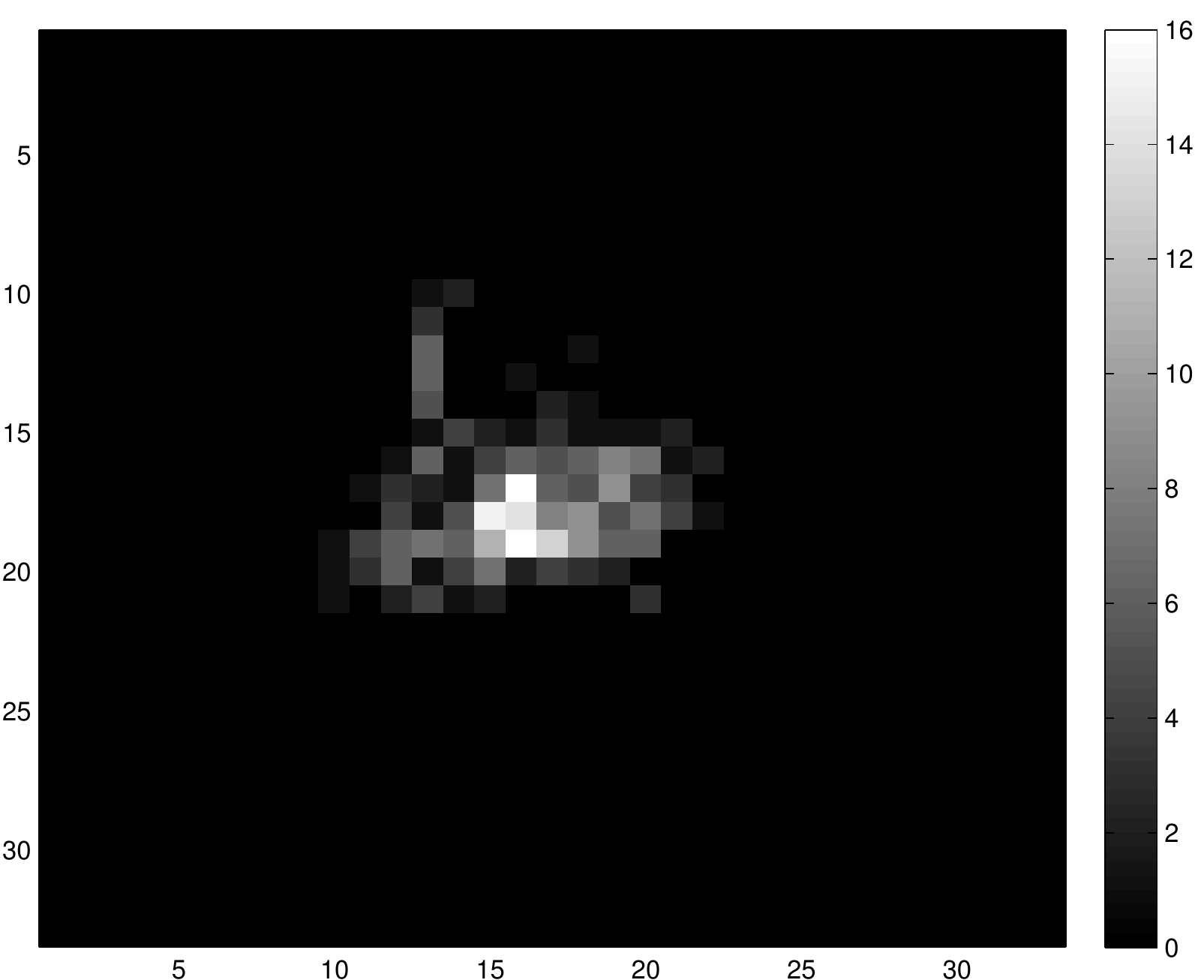}}
\subfigure[\label{centroids02} Centroid's occurrences for the \emph{QGoL}, for $\tau=0.1$ and $\sigma=0.5$.]
{\includegraphics[width=0.48\textwidth]{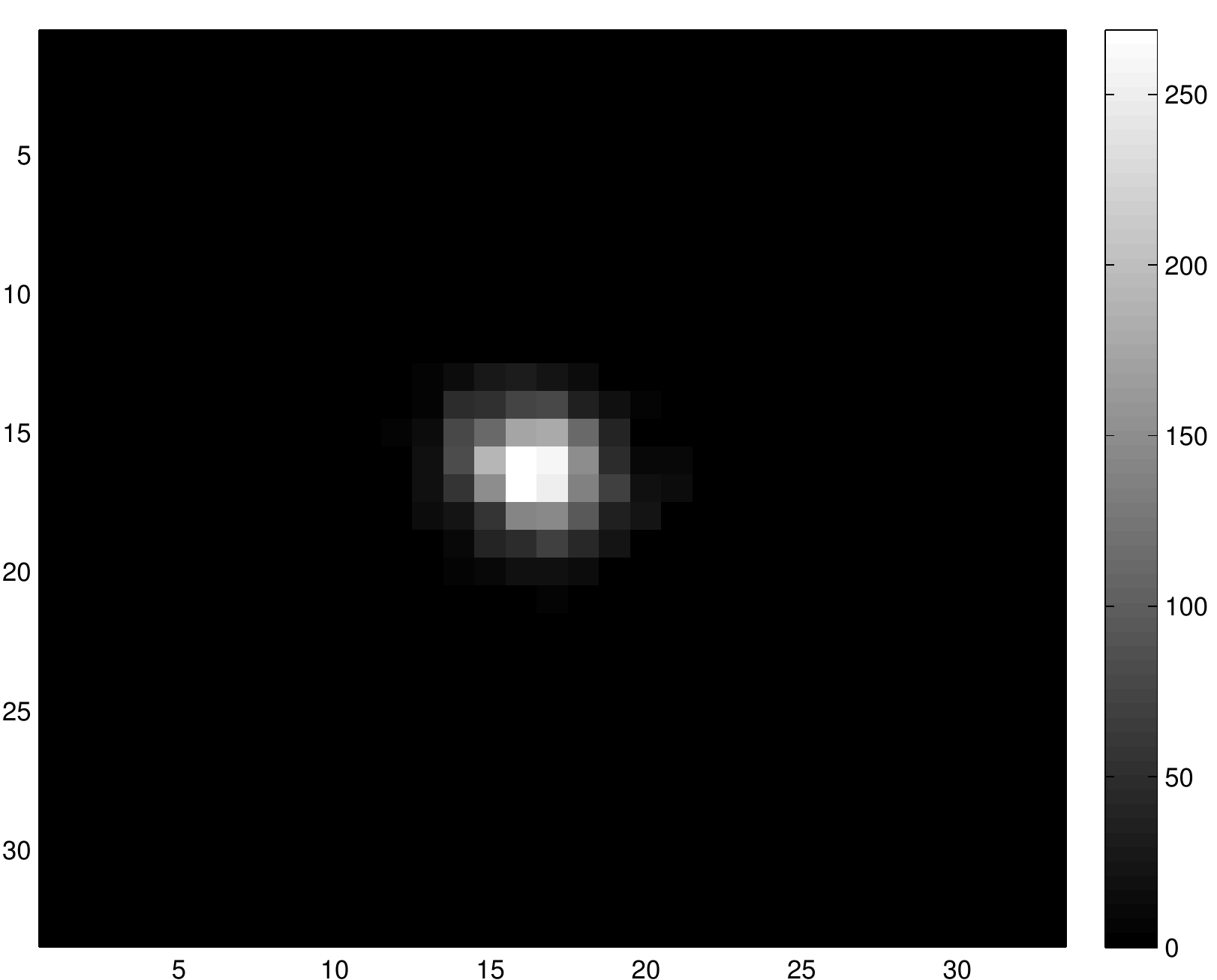}}
\end{center}
\caption{\footnotesize Number of occurrences of the center of mass of the whole system after each generation of the \emph{GoL}, \ref{centroids01}, and the \emph{QGoL} for parameters $\tau=0.1$ and $\sigma=0.5$, \ref{centroids02}, in the various cells of the lattice. Classical evolutions are generally characterized by the arrangement of the highest frequencies in an irregular, not centrally localized area. For quantum systems stabilizing after many generations (such as the case considered in \ref{centroids02}), instead, centroid occurrences appear enclosed in a narrow area composed of few centralized pixels.}
\end{figure}

Concerning the analysis of the centroids, the frequencies of the  occurrence of the center of mass of the whole binary images in the various cells of the lattice at each step of the classical and the quantum evolution have been analyzed. The study performed for successive generations shows that, as depicted in Figs.~\ref{centroids01}--\ref{centroids02}, while for the \emph{GoL} the highest frequencies are arranged in a fairly broad, irregular and not always centered area, for quantum games evolving for long times before reaching the stability we observe a shrinkage of this region to a distribution area with few centralized pixels.

\begin{figure}[h]
\begin{center}
\subfigure[\label{corr01} Sample correlation of blobs' centroids for the \emph{GoL}.]
{\includegraphics[width=0.48\textwidth]{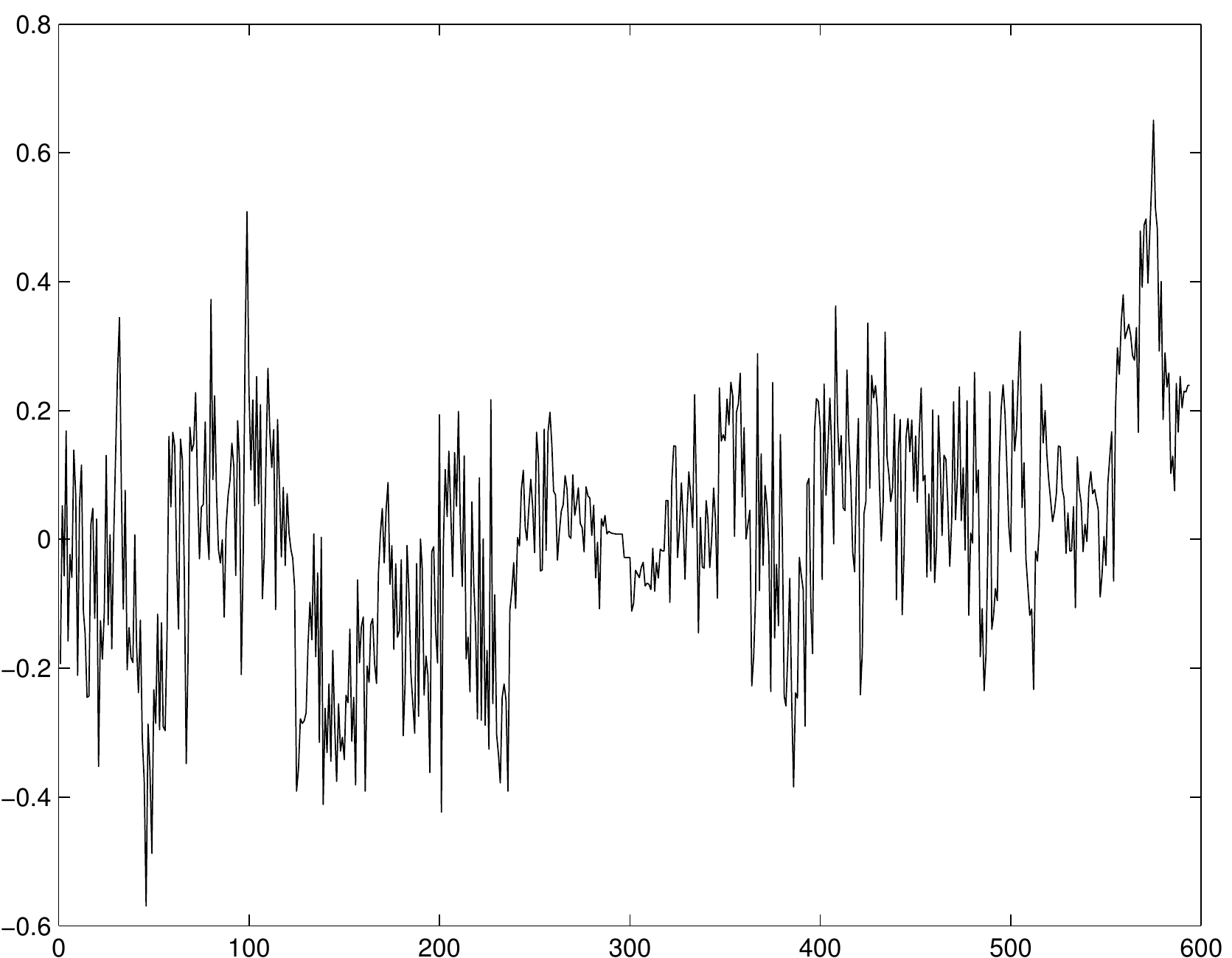}}
\subfigure[\label{corr02} Sample correlation of blobs' centroids for the \emph{QGoL}, for $\tau=0.25$ and $\sigma=1$.]
{\includegraphics[width=0.48\textwidth]{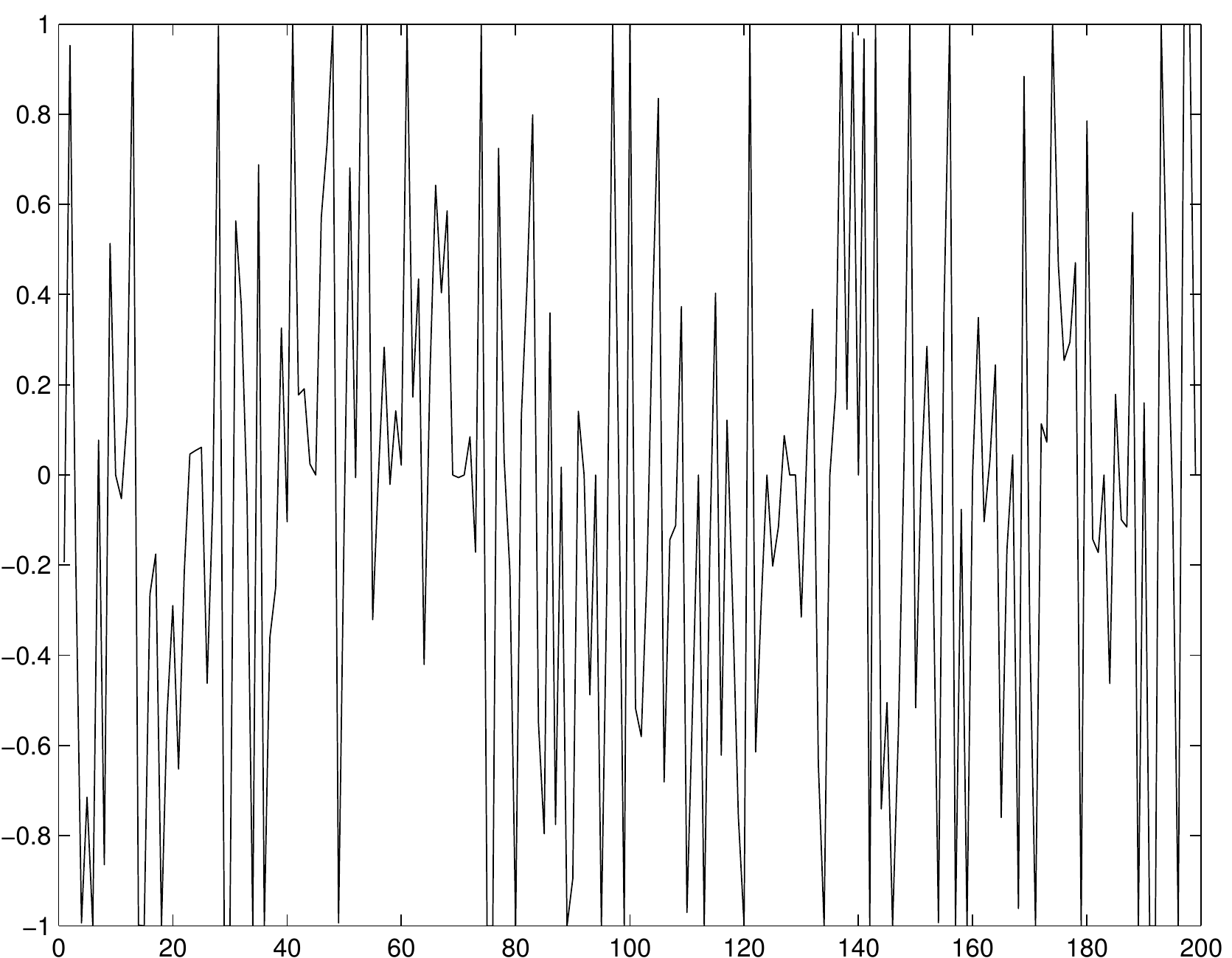}}
\end{center}
\caption{\footnotesize Sample correlation coefficients of the centroids of the various connected regions at each generation of the \emph{GoL}, \ref{corr01}, and the \emph{QGoL}, \ref{corr02}, for parameters $\tau=0.25$ and $\sigma=1$. In the case of quantum systems with $\tau$ greater than or equal to 0.25, the centroids of the blobs tend to assume configurations with direct or inverse correlation.}
\end{figure}

Moreover, the sample correlation coefficient of the cluster corresponding to the centroids of the connected regions at every generation has been taken into account. The trends recorded for values of $\tau$ greater than or equal to 0.25, associated with significant expected variations from the classical situation, highlight the fact that, as shown in Figs.~\ref{corr01}--\ref{corr02}, to a lack of sample correlation between the centers of mass of the blobs in the case of the classic game of life corresponds, rather, in the quantum setting, a tendency of these centroids to be arranged in configurations with direct or inverse correlation.

\section{Conclusions}\label{sect5}
In this paper we have discussed the possibility of introducing the time evolution of some macroscopic system by using some tools arising from quantum mechanics together with some specific rules. This is useful when, for instance, the time evolution of a given system $\Sc$ is driven not only by some Hamiltonian operator, as it happens for conservative closed microscopic systems or even for non-conservative microscopic open systems, but also by some external/internal action periodically applied to the system, and not easily included in any Hamiltonian. As we have seen, the rule can be seen as a sort of generalized projection operator, and its action on $\Sc$ may change  the original behavior of $\Sc$.

We have applied our idea to the \emph{Game of Life}, producing what we have called (our version of) the {\em quantum Game of Life}. A detailed analysis of this new system has been performed, and in particular we have discussed the role of the main parameters appearing in the model, and how their values affect the behavior of the system itself in comparison with what happens in the \emph{GoL}. More specifically, we have discussed how the \emph{QGol} is influenced by the transient time $\tau$ in which the quantum evolution governed by the Schr\"odinger equation takes place, and by the parameter $\sigma$ which modifies the classical rule adopted in the \emph{GoL}. We have found, through both the spectral and the blob analysis, that the \emph{QGoL} contains some really different evolutions with respect the \emph{GoL}, especially for moderate-high variations of the parameters $\tau$ and $\sigma$.

In our opinion, the definition of $(H,\rho)$--induced dynamics may open several possible lines of research, either from a theoretical viewpoint or in view of concrete applications, and we plan to apply this method to other concrete situations involving the same operational settings used in the situation considered in this paper.

\section*{Appendix: A case study: $L=5$}
\label{sec:L5}
Here, we consider a case study of \emph{GoL} and \emph{QGoL} dynamics choosing $L=5$. Since
in this case we have $2^{25}$ possible initial conditions, we may perform in a reasonable time a
complete analysis of all scenarios that can arise in the classical \emph{GoL} and how they differ from their quantum version. As remarked previously, for all the initial  conditions, $\phi^l_{\mathbf{n}},l=0,\ldots,2^{25}-1$,  a periodic behavior, in the sense of Definition~\ref{def2}, will emerge. Each initial condition (a distribution of 0 and 1 in the 25 cells of the lattice) can be considered as the binary representation of an integer in the range $[0,2^{25}-1]$; therefore, we may label each initial condition with the corresponding integer.

For $\tau=\sigma=0$, \emph{i.e.}, in the classical \emph{GoL}, following the evolution for all initial conditions, we have obtained that the possible evolutions lead to periodic solutions: the observed periods are 1, 2, 3, 4, 5, 10, and 20.
The data so obtained show that many initial conditions lead to the same periodic solutions, so that we can group the initial conditions in equivalence classes, see Table~\ref{table1}.

\begin{table}
\begin{center}
\begin{tabular}{c|ccccccc}
\hline
Period $\Omega$  & 1 & 2 & 3 & 4 & 5 & 10  & 20 \\
$\#$ of initial conditions & 3455 & 1225 & 200 & 200 & 20  & 60 &20\\
\hline
\end{tabular}
\end{center}
\caption{\label{table1}\footnotesize List all the possible period $\Omega$ for the \emph{Gol} in the case $L=5$, and number of equivalence classes of initial conditions having $\Omega$-periodic solutions.}
\end{table}

It is also interesting to consider how long the transient is for the different periodic solutions. For the initial conditions leading to the 1-equilibrium cyclic solutions, the length of transients  has a mean value of about 8;
most initial configurations have  very short transients (less than 7 generations), and, as the length of transients increases (its maximum is 51), the number of the initial conditions admitting them decays exponentially. For initial conditions leading to the
2--equilibrium cyclic solutions, the situation is quite the same:
the  length of transients  has a mean value of about 5;
most initial conditions have  very short transient (less than 5 generations), and, as the length
of transients increases (its maximum is 23), the number of the initial conditions admitting them
decays exponentially. For initial configurations leading to the 3--equilibrium cyclic solutions, the  length of transients   is in the range 1 to 4, and most of the initial conditions have one or two transient generations.

For initial conditions leading to the 4--equilibrium cyclic solutions, the
length of transients  has a mean value of about 8; most of the initial conditions  have a transient length between 1 and 13, and the maximum value  is 32. For initial conditions leading to  the 5--equilibrium cyclic solutions,
the length of transients  has a mean value of about 5, which is also the value with
the highest frequency; most of the remaining initial conditions exhibit almost uniformly distributed
transients of length equal to 1, 6, 7 and 8 (which is the maximum).
For initial conditions leading to the 10--equilibrium cyclic solutions, the length of transients  is in the range 1 to 10, and the distribution is almost uniform except for the extrema of the interval. Finally, for initial conditions leading to  the 20--equilibrium cyclic solutions,
the length of transients  has a mean value of about 4, and most of the initial conditions
have a transient  length between 1 and 2, while the maximum number of transients  is 15.

\begin{figure}[h]
\begin{center}
\subfigure[$\mathcal T_{3}(\tau,\sigma)$]{\includegraphics[width=0.48\textwidth]{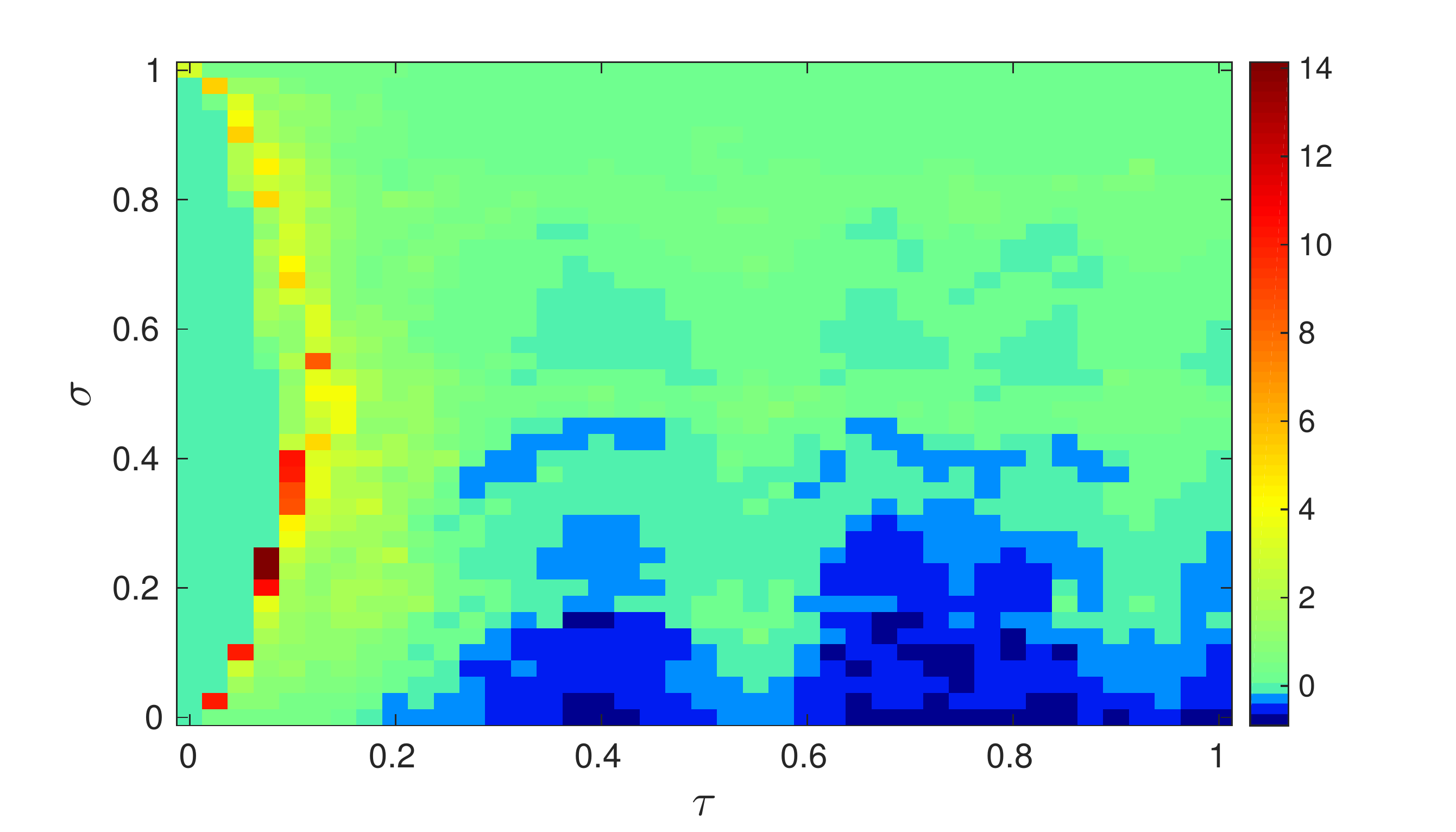} }\subfigure[$\Omega_{3}(\tau,\sigma)$]{\includegraphics[width=0.48\textwidth]{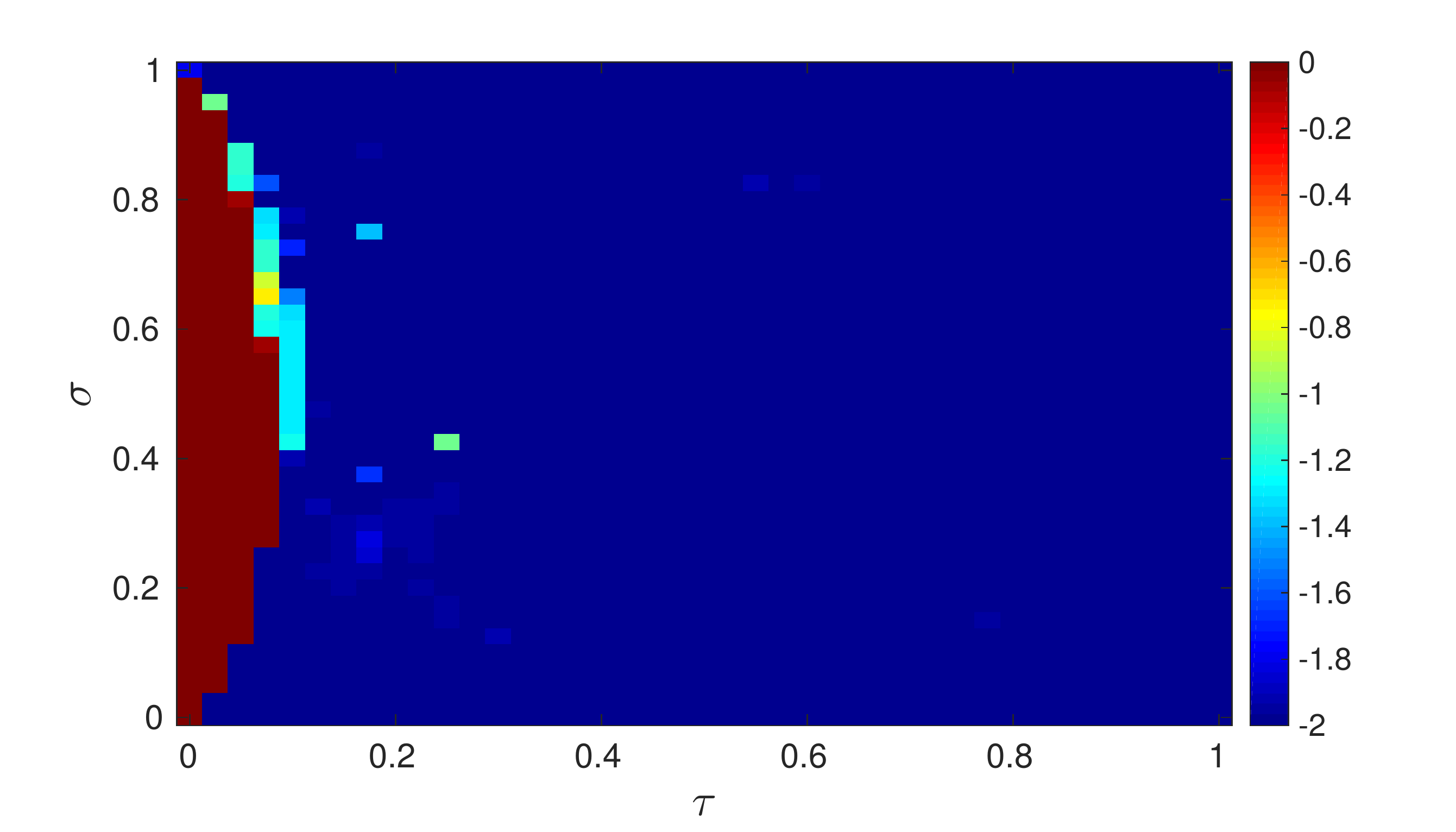}	}
\end{center}
\caption{\label{5x5I3}
\footnotesize The distribution $\mathcal T_P(\tau,\sigma)$ in \textbf{(a)}, and $\Omega_P(\tau,\sigma)$ in \textbf{(b)} for the period $P=3$.
For $\tau<\mathcal{C}(\sigma)$,  $\mathcal T_P(\tau,\sigma)$ and $\Omega_P(\tau,\sigma)$ vanish, hence that for $\tau<\mathcal{C}(\sigma)$ there is no substantial difference between the \emph{QGoL} and the \emph{GoL} case. The most relevant differences arise for  $\tau\approx\mathcal{C}(\sigma)$ where it is evident that
in the \emph{QGoL} case the solution is an equilibrium cyclic solution of period lower than the \emph{GoL} case and the transient to arrive to this solution is higher than
that in the \emph{GoL} case. $\tau>\mathcal{C}(\sigma)$ the situation is quite different, as periodicity of the \emph{QGoL} solution is in general lower than the \emph{Gol} case, and the transient can decrease or increase with respect the \emph{GoL} according to the various values of $\tau$ and $\sigma$.
}	
\end{figure}

\begin{figure}[!]
\begin{center}
\subfigure[$\mathcal T_{5}(\tau,\sigma)$]{\includegraphics[width=0.48\textwidth]{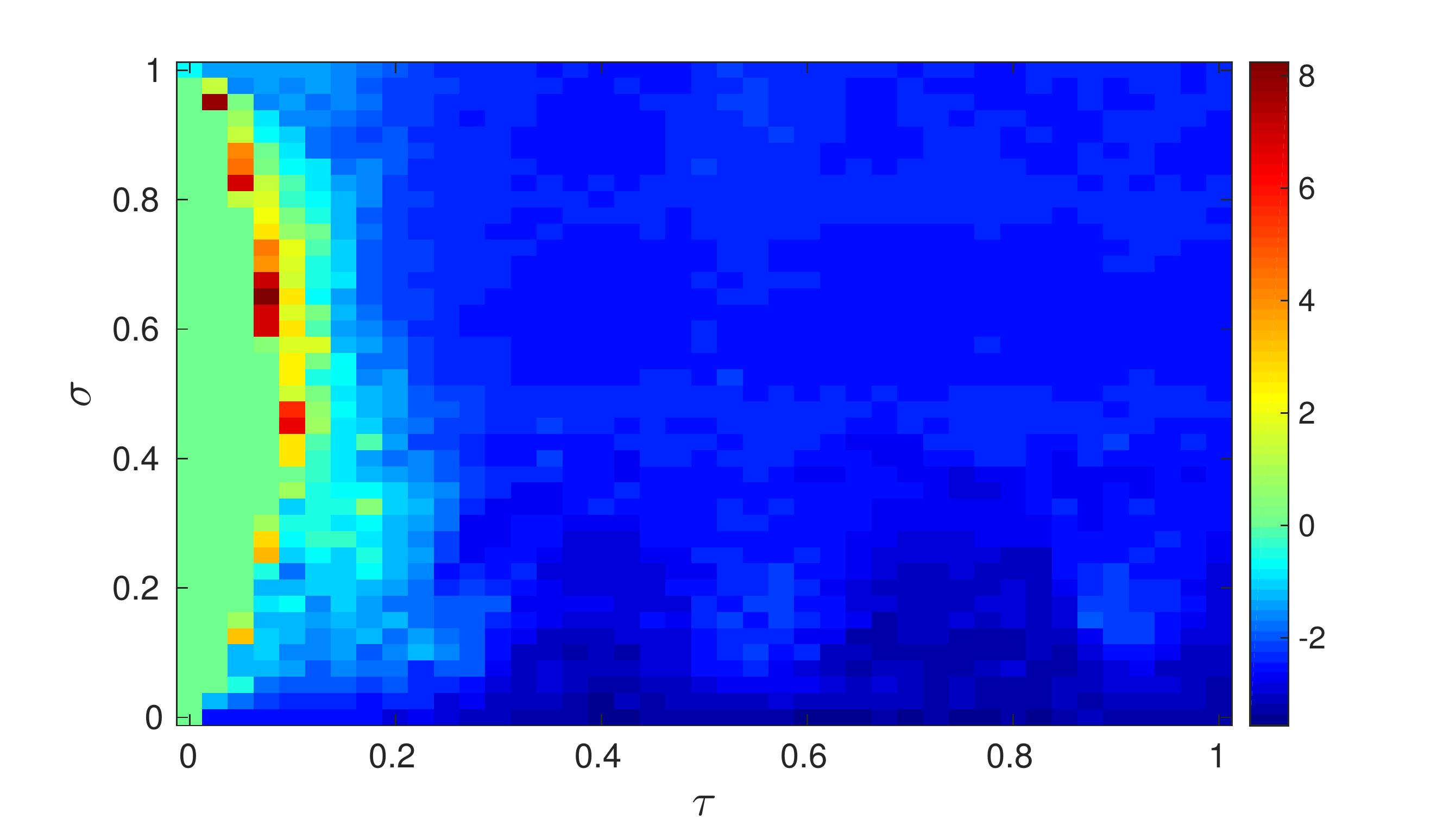} }
\subfigure[$\Omega_{5}(\tau,\sigma)$]{\includegraphics[width=0.48\textwidth]{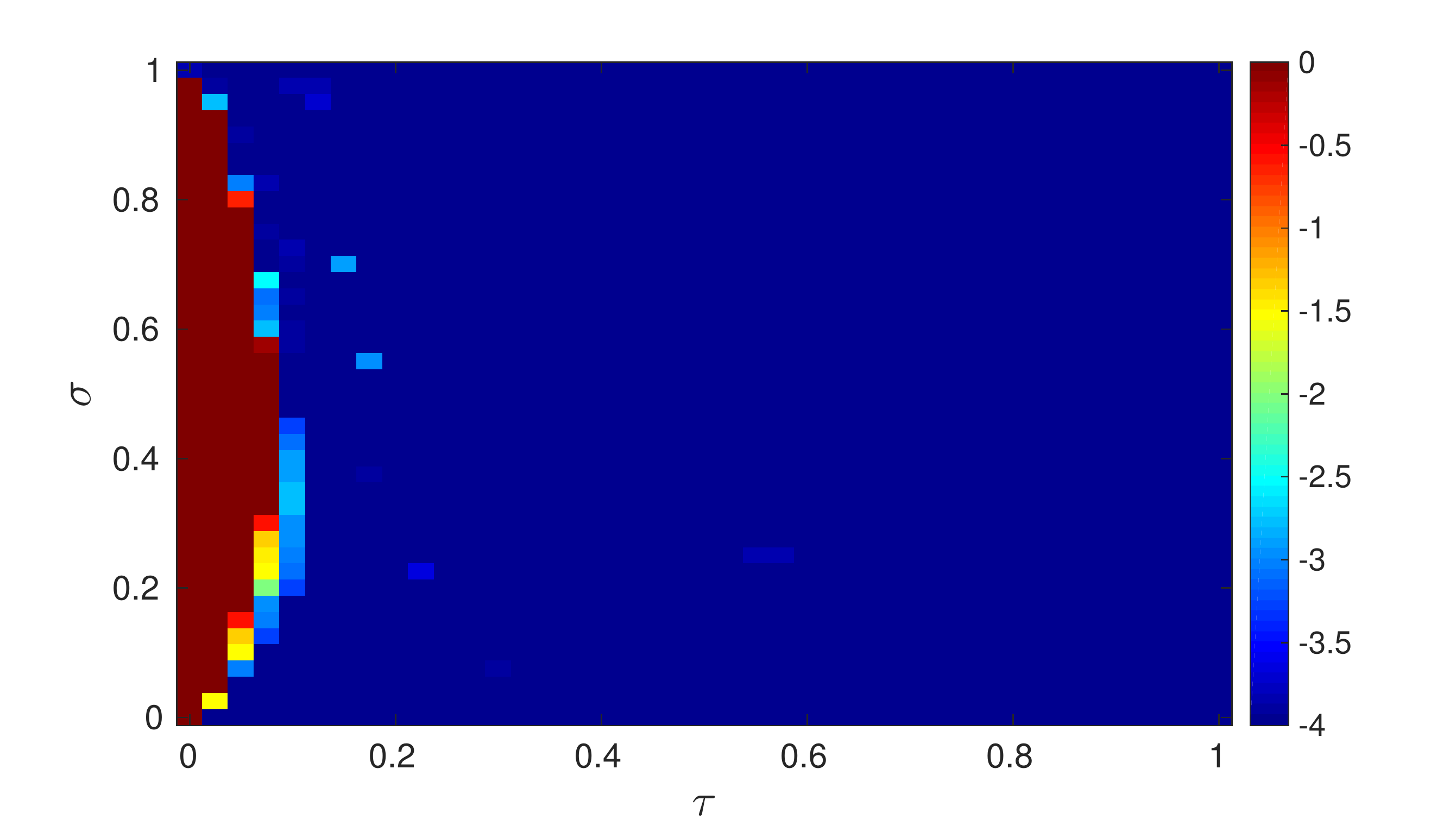}	}
\end{center}
\caption{\label{5x5I5}\footnotesize Same as Fig.~\ref{5x5I3} for $P=5$. Also in this case the main differences
between the \emph{GoL} and the \emph{QGoL} are visible for $\tau\approx\mathcal{C}(\sigma)$, and with
respect to the case $P=3$ the transient of the equilibrium cycle solution in the \emph{QGoL} case is alway
lower than the \emph{GoL} case.}	
\end{figure}

\begin{figure}[h]
\begin{center}
\subfigure[$\mathcal T_{10}(\tau,\sigma)$]{\includegraphics[width=0.48\textwidth]{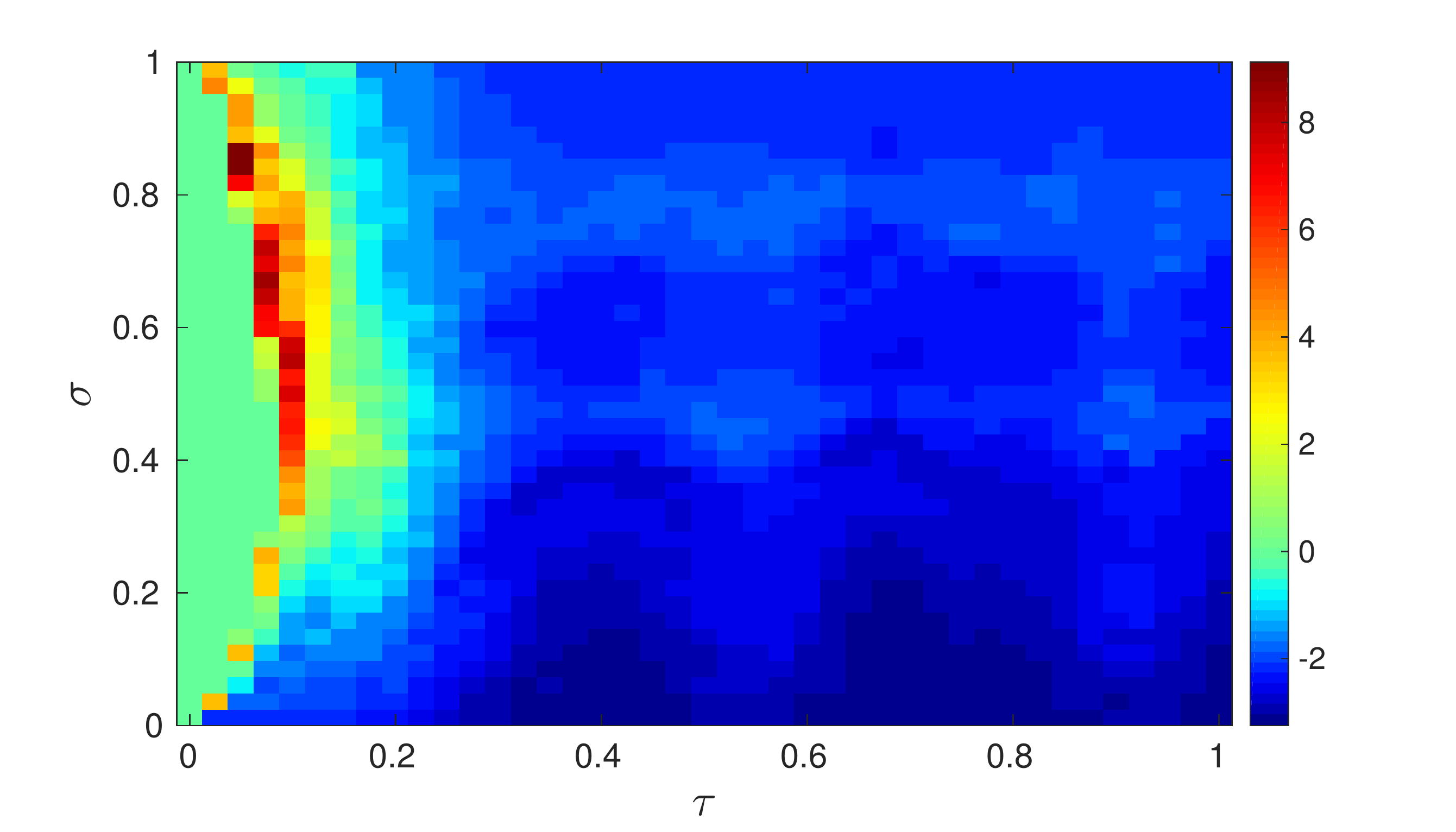} }	
\subfigure[$ \Omega_{10}(\tau,\sigma)$]{\includegraphics[width=0.48\textwidth]{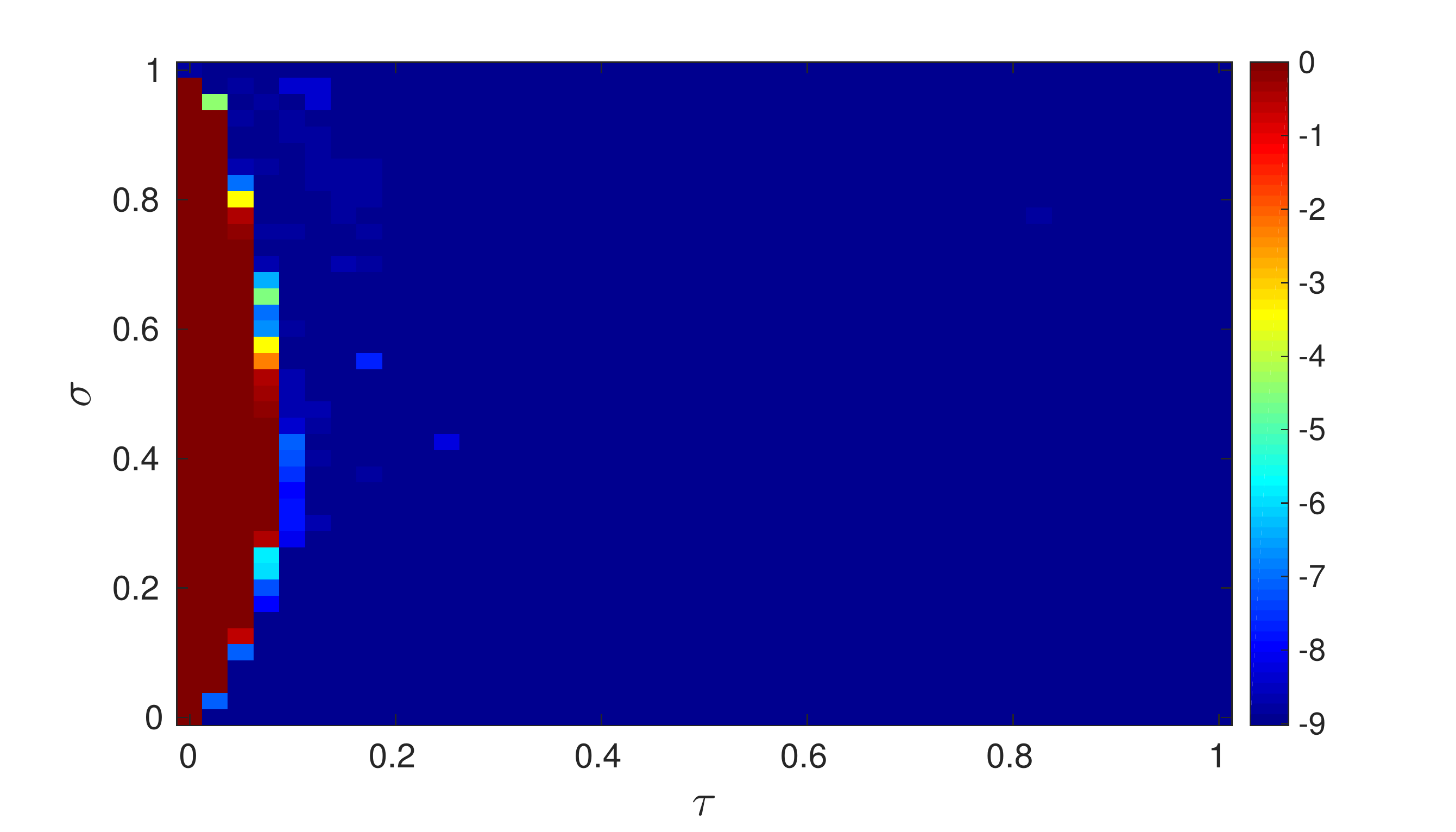}	}
\end{center}
\caption{\label{5x5I10}
\footnotesize Same as Fig.~\ref{5x5I3} for $P=10$. Results are similar to the case $P=5$.}	
\end{figure}

\begin{figure}[h]
\begin{center}
\subfigure[$\mathcal T_{20}(\tau,\sigma)$]{\includegraphics[width=0.48\textwidth]{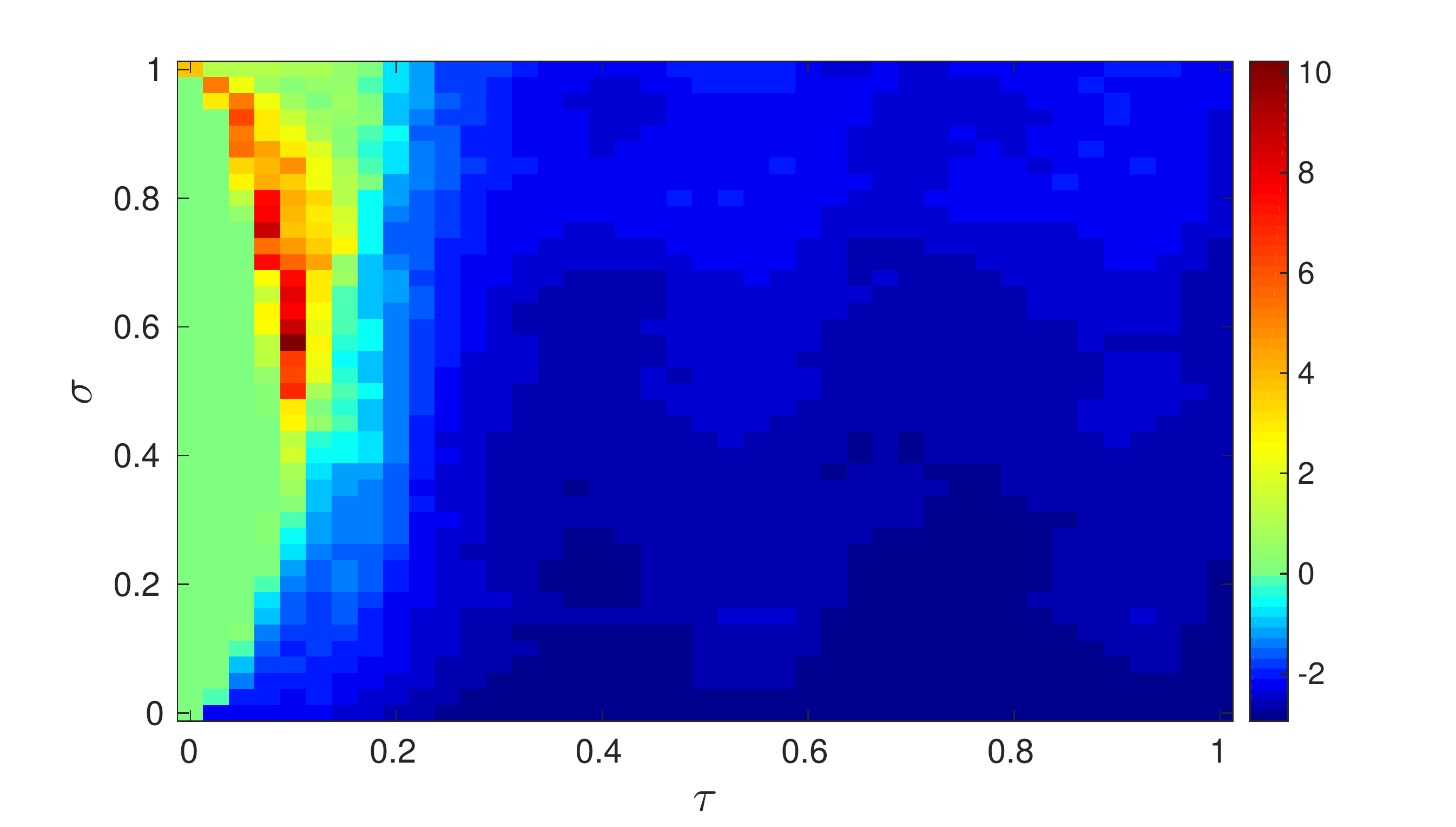} }	
\subfigure[$\Omega_{20}(\tau,\sigma)$]{\includegraphics[width=0.48\textwidth]{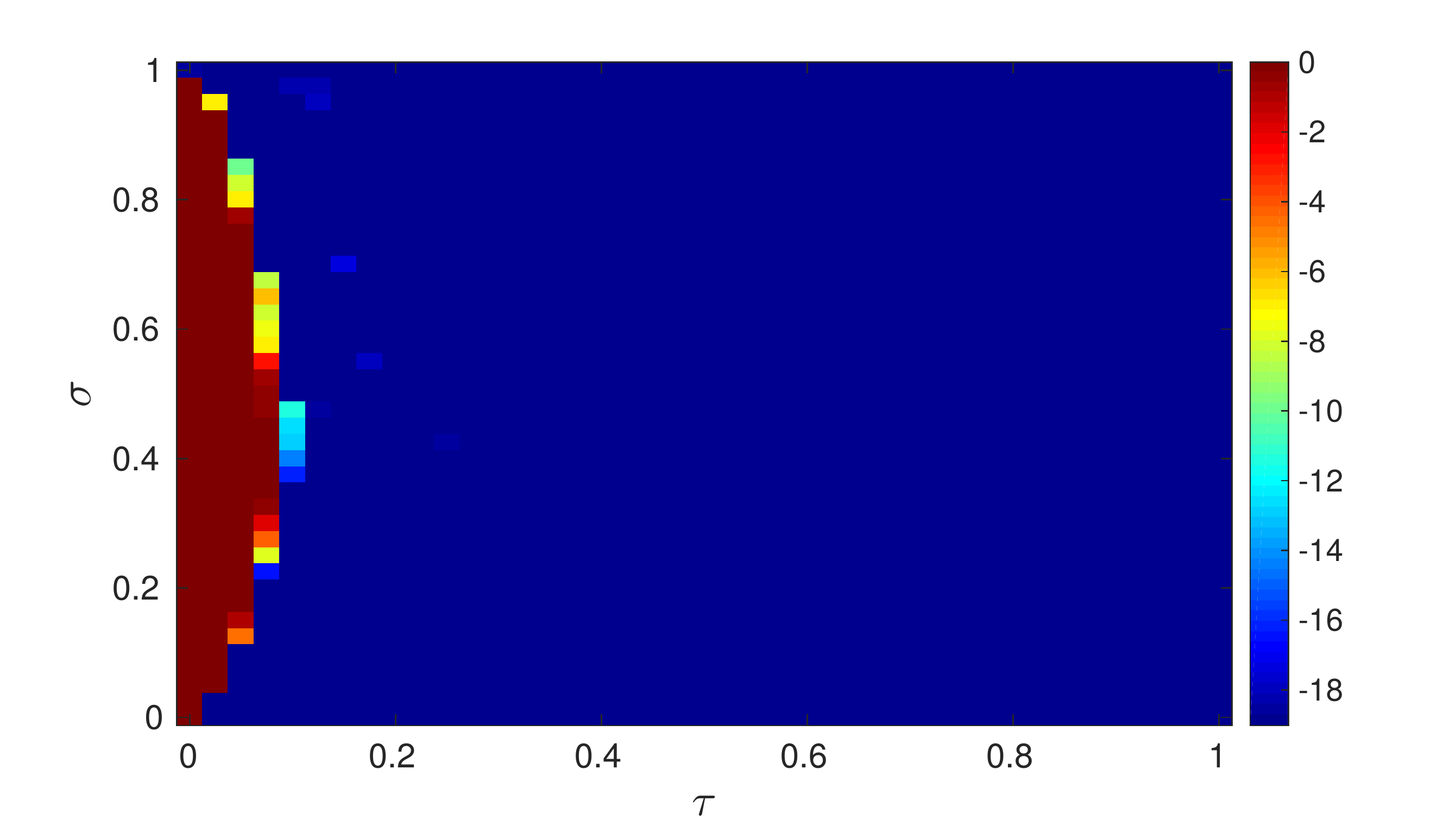}	}
\end{center}
\caption{\label{5x5I20}
Same as Fig.~\ref{5x5I3} for $P=20$. Results are similar to the case $P=5,10$.}	
\end{figure}

Consider now the behavior exhibited by the \emph{QGoL}, and let $T^\ell(\tau,\sigma)$ be the number of transient generations needed to reach a $P^\ell(\tau,\sigma)$-periodic solution  for a generic initial condition labeled with $\ell$. In the case of classical \emph{GoL} evolution, $\tau=0$, $\sigma=0$,  we have obtained the values of $P^\ell(0,0)$ (1, 2, 3, 4, 5, 10, 20, respectively). To investigate  how $T^\ell(\tau,\sigma)$ and $P^\ell(\tau,\sigma)$ are affected in the $(H,\rho)$--dynamics by the parameters $\tau$, $\sigma$, we  compute  the following mean distributions
\begin{eqnarray}
\mathcal T_P(\tau,\sigma)&=&\frac{1}{N_P}\sum_{k_P=1}^{N_P}\left(T^{j_{k_P}}(\tau,\sigma)-T^{j_{k_P}}(0,0)\right),\\
\Omega_P(\tau,\sigma)&=&\frac{1}{N_P}\sum_{k_P=1}^{N_P}\left(P^{j_{k_P}}(\tau,\sigma)-P^{j_{k_P}}(0,0)\right),
\end{eqnarray}
where $j_1,j_2,\ldots,j_{N_P}$ label the initial conditions having a period $P$. $\mathcal T_P(\tau,\sigma)$ and $\Omega_P(\tau,\sigma)$ allow to determine where the transient  and the periodic orbit length of the equilibrium cycle solution change according to the parameter $\tau$ and $\sigma$ with respect to the \emph{GoL}, as they are a measure of the variations between the \emph{GoL} and the \emph{QGoL} case. The results are shown in Figs.~\ref{5x5I3}-\ref{5x5I20} for the periodic orbit lengths $P=3,5,10,20$.  We can see that in terms of the transient length, the most relevant differences arises for $\tau=0.1,\sigma>0.5$ along the curve $\tau=\mathcal{C}(\sigma)=-0.337\sigma^2+0.384\sigma$ for $P=5,10,20$, while for $P=3$
the peaks are reached for $\tau<0.1,\sigma<0.5$ again along the curve $\tau\approx\mathcal{C}(\sigma)$.

For $\tau<\mathcal{C}(\sigma)$,  $\mathcal T_P(\tau,\sigma)$ and $\Omega_P(\tau,\sigma)$ vanish, meaning that for $\tau<\mathcal{\sigma}$ there is no substantial difference between the \emph{QGoL} and the \emph{GoL} case. This was already remarked in Section~\ref{sec:PTS}, where we noticed that   for $\tau<\mathcal{C}(\sigma)$  the distribution  $\Delta_{QGoL}^{GoL}(\tau,\sigma)$ has its lowest value. On the other hand, for $\tau>\mathcal{C}(\sigma)$ we obtain that the length of the periodic orbit $P^l(\tau,\sigma)$ is dramatically lower than the \emph{GoL} case.

\clearpage

\begin{thebibliography}{99}

\bibitem{AG} P. Arrighi, J. Grattage, Proc. JAC 201, Journes Automates Cellulaires 2010, Finland, 2010.

\bibitem{bagbook} F. Bagarello, {\em Quantum dynamics for classical systems: with applications of the
Number operator}, J. Wiley and Sons, New York, 2012.

\bibitem{BCO16} F. Bagarello, A.M. Cherubini, F. Oliveri, {\em An Operatorial Description of Desertification}, SIAM J. Appl. Math., 76(2), 479-499, 2016.


\bibitem{fff} F. Bagarello, F. Gargano, F. Oliveri, {\em A phenomenological operator description of dynamics of crowds: escape strategies},  Appl. Math. Model., \textbf{39}, 2276--2294, 2015.

\bibitem{ff2} F. Bagarello, F. Oliveri, {\em An operator description of interactions between populations with applications to migration}, Math. Mod. Methods Appl. Sci. \textbf{23},  471--492, 2013.

\bibitem{BCM2012} D. Bleh, T. Calarco, S. Montagero, {\em Quantum Game of Life}, EPL A Letters Journal Exploring the Frontiers of Physics, \textbf{97}:20012, 2012.

\bibitem{Deu} D. Deutsch, {\em Quantum theory, the Church-Turing principle and the universal quantum computer}, Proc.  Royal Society of London A, \textbf{400}, 97--117, 1985.

\bibitem{DO16} R. Di Salvo, F. Oliveri, {\em An operatorial model for long-term survival of bacterial populations}, Ricerche di Matematica, doi:10.1007/s11587-016-0266-z, 1-13, 2016.

\bibitem{Fey} R. Feynman, {\em Simulating physics with computers}, Int. J. Theor. Phys., \textbf{21},  467--488, 1982.

\bibitem{FA} A.P. Flitney, D. Abbott, {\em Towards a Quantum Game of Life}, Game of Life Cellular Automata, Springer London, 465--486, 2010.

\bibitem{gar} F. Gargano, {\em Dynamics of Confined Crowd Modelled Using Fermionic Operators}, Int. J. Th. Phys. \textbf{53}, 2727--2738, 2014.

\bibitem{GZ} G. Grossing, A. Zeilinger, {\em Structures in quantum cellular automata}, Phys. B, \textbf{151}, 366--370, 1988.

\bibitem{qdm2} E. Haven, A. Khrennikov, \emph{Quantum social science},
Cambridge University Press, New York, 2013.

\bibitem{kes} M.S. Keshner, {\em 1/f Noise}, Proceedings of the IEEE, \textbf{70},  212--218, 1982.

\bibitem{qdm1} A. Khrennikov, {\em Ubiquitous quantum structure: from psychology to finances}, Springer, Berlin, 2010.

\bibitem{LAPM} J. Lee, S. Adachi, F. Peper, K. Morita, {\em Asynchronous game of life}, Physica D.,\textbf{194}, 369--384, 2004.

\bibitem{blob} T.~Lindeberg, {\em Scale-Space Theory in Computer Vision}, Springer, 1994.

\bibitem{mer} E. Merzbacher, {\em Quantum Mechanics}, John Wiley and Sons, New York, 1998.

\bibitem{mes} A. Messiah, {\em Quantum mechanics}, North Holland Publishing Company, Amsterdam, 1961.

\bibitem{NYH} S. Ninagawa, M. Yoneda, S. Hirose, {\em $1/f$ fluctuation in the ``Game of Life''}, Physica D, \textbf{118}, 49--52, 1998.

\bibitem{NYH2} S. Ninagawa, {\em Power Spectral Analysis of Elementary Cellular Automata}, Complex Systems, \textbf{17}, 399--411, 2008.

\bibitem{rs} M. Reed, B. Simon, {\em Methods of Modern Mathematical Physics}, I, Academic Press, New York, 1980.

\bibitem{roman}P. Roman, {\em Advanced quantum mechanics}, Addison--Wesley, New York, 1965.



\end{thebibliography}
\end{document}